\documentclass[12pt]{article}
\usepackage{amsmath}
\usepackage{amssymb,amsthm}
\usepackage[english]{babel}
\usepackage[textwidth=18.5cm,textheight=22cm]{geometry}
\usepackage{graphicx}
\usepackage{psfrag}
\newcommand{\C}{\mathbb{C}}

\newcommand{\R}{\mathbb{R}}
\newcommand{\Z}{\mathbb{Z}}
\newcommand{\N}{\mathbb{N}}

\newcommand{\bM}{\mathcal{M}}
\newcommand{\bZ}{\mathcal{Z}}

\newcommand{\A}{\mathcal{A}}
\renewcommand{\d}{\mathrm{d}}
\newcommand{\bt}{\boldsymbol{t}}

\newcommand{\bA}{\boldsymbol{A}}
\newcommand{\bu}{\boldsymbol{u}}
\newcommand{\bv}{\boldsymbol{v}}

\newcommand{\be}{\boldsymbol{e}}
\newcommand{\bc}{\boldsymbol{c}}
\newcommand{\bR}{\boldsymbol{R}}
\newcommand{\bs}{\boldsymbol{s}}
\newcommand{\bn}{{\boldsymbol{n}}}

\newtheorem{pro}{Proposition}

\newtheorem{teh}{Theorem}

\begin{document}

\title{ Multiple orthogonal polynomials, string equations\\ and the large-$\bn$ limit
\thanks{Partially supported by  MEC
project FIS2008-00200/FIS and ESF programme MISGAM}}
\author{L. Mart\'{\i}nez Alonso$^{1}$ and E. Medina$^{2}$
\\
\emph{$^1$ Departamento de F\'{\i}sica Te\'{o}rica II, Universidad
Complutense}\\ \emph{E28040 Madrid, Spain}\\
\emph{$^2$ Departamento de Matem\'aticas, Universidad de C\'adiz}\\
\emph{E11510 Puerto Real, C\'adiz, Spain} }

\date{} \maketitle

\abstract{}

The Riemann-Hilbert problems for multiple orthogonal polynomials of
types I and II are used to derive 
string equations associated to pairs of Lax-Orlov operators.  A method for determining the quasiclassical limit of
string equations in the phase space of the  Whitham
hierarchy of dispersionless integrable systems is provided.  Applications to the analysis of the large-$\bn$ limit of multiple orthogonal polynomials and their  associated random
matrix ensembles and models of non-intersecting Brownian motions are given.

\vspace*{.5cm}

\begin{center}\begin{minipage}{12cm}
\emph{Key words:} Multiple orthogonal polynomials. String equations.  Whitham hierarchies.
 \emph{PACS number:} 02.30.Ik.
\end{minipage}
\end{center}
\newpage

\section{Introduction}

The set of orthogonal
polynomials
$
P_n(x)=x^n+\cdots,
$
with respect to an exponential
weight
\[
\int_{-\infty}^{\infty}P_n(x)\,P_m(x)\,e^{V(\bc,x)}\,
d x=h_n\delta_{nm},\quad V(\bc,x):=\sum_{k\geq 1}c_k\,x^k,
\]
is an essential ingredient of the methods \cite{gin}-\cite{dei} for studying the large-$n$ limit of the Hermitian matrix model
\begin{equation}\label{i2}
Z_n=\int \d M \exp\Big(\mbox{Tr}\,V(\bc,M)\Big).
\end{equation}
One of the  main tools used in these methods is the pair of equations
\begin{equation}\label{add}
z\,P_n(z)=\bZ\,P_n(z),\quad
\partial_z\,P_n(z)=\bM\,P_n(z),\quad n\geq0,
\end{equation}
where $(\bZ,\bM)$ is a pair of Lax-Orlov operators of the form
\begin{equation}\label{diss}
\bZ=\Lambda+u_n+v_n\,\Lambda^*,\quad \bM=-\sum_{k\geq
1}k\,c_k\,(\bZ^{k-1})_+.
\end{equation}
Here $\Lambda$ is the shift matrix acting in the linear space of
sequences,
 $\Lambda^*$ is its transposed matrix and $(\quad)_+$ denotes the lower part (below the main diagonal) of semi-infinite matrices.

\vspace{0.3cm}

  The first equation in \eqref{add} represents the standard three-term relation for orthogonal polynomials. Both equations are referred to as the \emph{string equations} in  the matrix models of 2D quantum gravity \cite{gin} and
provide the starting point of several techniques to characterize   the large-$n$ limit of \eqref{i2}.
A deeper mathematical insight of these methods was achieved  after the introduction by Fokas, Its and Kitaev \cite{fok} of a matrix valued Riemann-Hilbert (RH) problem which characterizes orthogonal polynomials on the real line, and the formulation by Deift and Zhou \cite{dei},\cite{dei2} of steepest descent methods for studying asymptotic
properties of RH problems .

\vspace{0.3cm}

The RH problem of Fokas-Its-Kitaev was generalized by Van Assche, Geronimo and Kuijlaars
\cite{jer} to characterize multiple orthogonal polynomials. Moreover, it was found \cite{ortb1}-\cite{ortb5}
that these families of polynomials are closely connected to important statistical models such as Gaussian ensembles with
external sources and one-dimensional non-intersecting Brownian motions.

\vspace{0.3cm}

In this paper we generalize the string equations \eqref{add} to
multiple orthogonal polynomials of types I and II, and show how
these equations can be applied to analyze the large-$\bn$ limit of
multiple orthogonal polynomials and their associated statistical
models.
Section 2 introduces  the basic strategy of our approach to derive
string equations, which is inspired by  standard methods used in
the theory of multi-component integrable systems \cite{krich}-\cite{mio00}. As it was proved in \cite{jer} the multiple orthogonal polynomials
of types I and II  are elements of the first row of the fundamental solution $f$ of the corresponding RH problem.  Then, in Sections 3 and 4 we formulate systems of string equations for the elements of the first row of the fundamental
 solution $f$.
 In both
cases the function $f$ depends on a set of discrete variables
\[
\bs=(s_1,s_2,\ldots,s_q)\in\Z^{q},\quad \mbox{where}\quad
\begin{cases}
\mbox{$s_i\geq 0$ for type I polynomials}\\
 \mbox{$s_i\leq 0$ for type II polynomials}.
\end{cases}
\]
 Therefore,
special care is required to determine the form of the string
equations  on the boundary of the domain of the discrete variables. Thus, we obtain
closed-form expressions, free of boundary terms, for the string equations
satisfied by these types of multiple orthogonal polynomials. These string equations are associated to pairs $(\bZ_i,\bM_i)$ of Lax-Orlov operators. In particular those involving the Lax operators $\bZ_i$ lead to the well-known recurrence relations for multiple orthogonal polynomials \cite{jer}.

\vspace{0.3cm}

We take advantage of an interesting observation due to
Takasaki and Takebe \cite{tak2} who showed that the
 dispersionless limit of a row of a matrix-valued KP wave function is a solution of
the zero genus Whitham hierarchy \cite{krich}. This is an additional incentive  for
using  Lax-Orlov operators
 \cite{tak1}-\cite{mio00} in order  to characterize the large-${\bn}$ limit in terms of quasiclassical (dispersionless limit)
expansions. Thus, in Section 5  we show how the leading term of the
expansion of the first row of $f$ is determined by a
system of dispersionless string equations for $q+1$ Lax-Orlov
functions $(z_{\alpha},m_{\alpha})$ in the phase space of the
Whitham hierarchy. The unknowns of this system reduce to a set of  $q$ pairs of
functions $(u_k,v_k)$, which are determined by means of a system of
hodograph type equations. Finally, Section 6 is devoted to
illustrate the applications of our results to models of random
matrix ensembles and non-intersecting Brownian motions.

\vspace{0.3cm}

 The present work deals with multiple orthogonal
polynomials of types I and II only, but the same
 considerations apply  to  the
study of multiple orthogonal polynomials of mixed type \cite{kuij}. On
the other hand, we concentrate on the description of the leading terms of
the  asymptotic solutions in the dispersionless
limit. However, as it was showed in \cite{mio1}-\cite{mio2} for the
case of the Toda hierarchy and the Hermitian matrix model, the
scheme used in the present paper can be further elaborated  for
determining the general terms of these expansions, as well as their
critical points and their corresponding double scaling limit
regularizations.

\section{Riemann-Hilbert problems}
In this work we will consider $(q+1)\times (q+1)$ matrix valued
functions.  Unless otherwise stated Greek
$\alpha,\beta,\ldots$ and Latin $i,j,\ldots$ suffixes will  label
indices of the sets $\{0,1,\ldots,q\}$ and $\{1,2,\ldots,q\}$,
respectively. We will denote by $E_{\alpha\beta}$  the matrices
$(E_{\alpha\beta})_{\alpha'\beta'}=\delta_{\alpha\alpha'}
\,\delta_{\beta\beta'}$ of the canonical basis and, in particular,
its diagonal members will be denoted by
$E_{\alpha}:=E_{\alpha\alpha}$. Some useful relations which will be
frequently used in the subsequent discussion are
\[
E_{\alpha\beta}\,E_{\gamma\lambda}=\delta_{\beta\gamma}\,E_{\alpha\lambda};\quad \quad
E_{\alpha}\,a\,E_{\beta}=a_{\alpha\beta}\, E_{\alpha\beta},\quad
\forall \;\mbox{matrix $a$}.
\]
We will also denote by $V(\bc,z)$ the scalar function
\begin{equation}\label{uve}
V(\bc,z):=\sum_{n\geq 1}c_n\,z^n,\quad
\bc=(c_{1},c_{2},\ldots)\in\C^{\infty},
\end{equation}
and will assume that only a finite number of the coefficients $c_n$ are different from zero.

\vspace{0.2cm}

Given a matrix function $g=g(z) \; (z\in \R)$
such that $\mbox{det}\,g(z)\equiv 1$, we will consider the RH problem
\begin{equation}\label{rhp}
m_-(z)\, g(z)=m_+(z),\quad z\in\R,
\end{equation}
where $m(z)$ is a sectionally holomorphic function and $m_{\pm}(z):=\lim_{\epsilon\rightarrow 0+}\,m(z\pm
i\epsilon)$.
We are interested in solutions $f=f(\bs,z)$ of \eqref{rhp} depending on $q$ discrete
variables $\bs=(s_1,\ldots,s_q)\in\Z^{q}$ such that
\begin{equation} \label{ae}
f(\bs,z)=\Big(I+\mathcal{O}\Big(\dfrac{1}{z}\Big)\Big)\,f_0(\bs,z),\quad
z\rightarrow\infty,
\end{equation}
where
\[
f_0(\bs,z):=\sum_{\alpha=0}^q z^{s_{\alpha}}\,E_{\alpha},\quad
(s_0:=- \sum_{i=1}^q\,s_i).
\]
The set of points $\bs\in \Z^q$
for which \eqref{rhp} admits a solution $f(\bs,z)$ satisfying
\eqref{ae} will be denoted by $\Gamma$. The solution  $f(\bs,z),\, (\bs\in\Gamma)$ is unique and will be referred to as the \emph{fundamental solution} of the RH problem \eqref{rhp}.

\vspace{0.2cm}

We will apply \eqref{rhp} and \eqref{ae} to derive    certain difference-differential equations for $f$. These  equations contain two basic ingredients: the coefficients of the asymptotic expansion of $f(\bs,z)$ as
$z\rightarrow\infty$
\begin{equation}\label{mata}
f(\bs,z)=\Big(I+\sum_{n\geq
1}\dfrac{a_n(\bs)}{z^n}\Big)\,f_0(\bs,z),
\end{equation}
and the $q$ pairs of shift operators $T_i,\,T^*_i$
acting on functions $h(\bs)\,(\bs\in\Gamma)$ defined as
\begin{equation*}
(T_i\,h)(\bs):=\begin{cases}
h(\bs-\be_i)\;\, \mbox{if $\bs-\be_i\in \Gamma$}\\\\
0 \;\; \mbox{if $\bs-\be_i\notin \Gamma$}
\end{cases},\quad
(T_i^*\,h)(\bs):=\begin{cases}
h(\bs+\be_i)\; \;\mbox{if $\bs+\be_i\in \Gamma$}\\\\
0\;\; \mbox{if $\bs+\be_i\notin \Gamma$},
\end{cases}
\end{equation*}
where $\be_i$ are the elements of the canonical
basis of $\C^q$.

\vspace{0.3cm}
 We will often consider series of the form
\[
\A:=\sum_{n=1}^{ \infty} c_n(\bs)\,(T_i^*)^n+c'_0+\sum_{n=1}^{
\infty} c'_n(\bs)\,T_i^n,
\]
and will denote
\begin{equation}\label{proy}
(\A)_{(i,+)}:=\sum_{n=1}^{ \infty} c_n(\bs)\,(T_i^*)^n,\quad
(\A)_{(i,-)}:=c'_0+\sum_{n=1}^{ \infty} c'_n(\bs)\,T_i^n.
\end{equation}

 \vspace{0.3cm}
The RH problem \eqref{rhp}
admits the following symmetries.

\begin{pro}

\begin{enumerate}
\item If $h(\bs,z)\,(\bs\in\Gamma)$ is an
 entire  function of $z$, then $h(\bs,z)\,f(\bs,z)$ satisfies \eqref{rhp} for all $\bs\in\Gamma$.
\item   The functions $(T_i\,f)(\bs,z)$ and $(T_i^*\,f)(\bs,z)$ satisfy \eqref{rhp} for all $\bs\in\Gamma$.

\item If $g(z)$ is an entire function, then for any entire  function $\phi(z)$ verifying
\begin{equation}\label{inv}
g^{-1}\,\phi\,g=\phi-g^{-1}\,\partial_z\, g,
\end{equation}
 the \emph{covariant derivative}
\begin{equation}\label{cov}
D_z\,f:=\partial_z \,f-f\,\phi,
\end{equation}
satisfies \eqref{rhp} for all $\bs\in\Gamma$.
\end{enumerate}
\end{pro}

\noindent

Our strategy to obtain   difference-differential equations for $f$ is based on applying the next simple statement to the symmetries of \eqref{rhp}.

\begin{pro} Let $\tilde{f}(\bs,z)$ be a a solution of \eqref{rhp}
defined for $\bs$ in a certain subset $\Gamma_0 \subset\Gamma$.
 If $\tilde{f}(\bs,z)\,f(\bs,z)^{-1}-P(\bs,z)\rightarrow 0$ as $z\rightarrow
\infty$, where $P(\bs,z)$ is a
polynomial in $z$, then 
\[ \tilde{f}(\bs,z)=P(\bs,z)\,f(\bs,z). 
\]
\end{pro}
\begin{proof}
Since $\mbox{det}\,g(z)\equiv 1$ it follows from \eqref{rhp} and
\eqref{ae} that $\mbox{det}\,f(\bs,z)\equiv 1$ so that the inverse matrix
 $f(\bs,z)^{-1}$ is analytic  for $z\in \C-\R$ and satisfies the jump condition
\begin{equation*}
g(z)^{-1}\,f_-(\bs,z)^{-1}\, =f_+(\bs,z)^{-1},\quad z\in\R.
\end{equation*}
As a consequence  $\tilde{f}\,f^{-1}$ is an entire function of $z$ and
the statements follow at once.
\end{proof}

\section{Multiple orthogonal polynomials of type I}

Given $q$ exponential weights $w_i$ on the real line 
\[
\quad w_i(x):=e^{-V(\bc_i,x)},\quad
\bc_i=(c_{i1},c_{i2},\ldots)\in\C^{\infty},
\]
and
$\bn=(n_1,\ldots,n_q)\in\N^q$ with $|\bn|\geq1$, the
type I orthogonal polynomials
\[
\bA(\bn,x)=(A_1(\bn,x),\dots,A_q(\bn,x))
\]
are determined by means of  the following
conditions:
\begin{description}
\item[i)] If $n_j\geq1$ then the polynomial  $A_j(\bn,x)$ has degree   $n_j-1$. If $n_j=0$ then $A_j(\bn,z)\equiv0$
\item[ii)] The following orthogonality relations are satisfied
$$\int_{\mathbb R}\frac{dx}{2\pi i}
x^l\left(\sum_{j=1}^qA_j(\bn,x)w_j(x)\right)=\left\{
\begin{array}{ll}
0 & l=0,1,\dots,|\bn|-2,\\
1 & l=|\bn|-1.
\end{array}\right.$$
\end{description}
We assume that all the multi-indices $\bn$ are strongly normal
\cite{kuij} so that $A_j(\bn,z)$ are unique.

\vspace{0.3cm}

The  RH problem which characterizes these polynomials \cite{jer} is determined by
\begin{equation}\label{gtypeI}
g(z)=\left(
\begin{array}{ccccc}
1&0&0&\ldots&0\\
-w_1(z)&1&0&\ldots&0\\
\vdots&\vdots&\vdots&\vdots\\
-w_q(z)&0&\ldots&0&1
\end{array}
\right).
\end{equation}
The corresponding fundamental solution $f(\bs,z)$ exists on the domain
\begin{equation}\label{d1}
\Gamma_{I}=\{\bs\in\Z^{q}\,:\, s_i\geq 0,\,\forall
i=1,\ldots,q\}.
\end{equation}
For $ \bs\neq \boldsymbol{0} $ it is given by
\begin{align}\label{sol1}
\nonumber f(\bs,z)&=\left(
\begin{array}{cc}
R(\bs,z) & \bA(\bs,z)\\
d_1^{-1}R(\bs+\be_1,z) & d_1^{-1}\bA(\bs+\be_1,z)\\
\vdots&\vdots\\
d_q^{-1}R(\bs+\be_q,z) & d_q^{-1}\bA(\bs+\be_q,z)
\end{array}
\right),\\\\
\nonumber R(\bs,z)&:=\int_{\mathbb{R}}\dfrac{\d x}{2\pi i}
\displaystyle\dfrac{\sum_{j=1}^qA_j(\bs,x)\,w_j(x)}{z-x},
\end{align}
where $d_j$ is the leading coefficient of  $A_j(\bs+e_j,z)$.
Furthermore, for $\bs=\boldsymbol{0}$
\begin{equation}\label{sol10}
f(\boldsymbol{0},z)=\left(
\begin{array}{ccccc}
1&0&0&\cdots&0\\
R_1(z)&1&0&\cdots&0\\
R_2(z)&0&1&\cdots&0\\
\vdots&\vdots&\vdots&  &\vdots\\
R_q(z)&0&0&\cdots&1
\end{array}
\right),\quad R_j(z):=\int_{\mathbb{R}}\dfrac{\d x}{2\pi i}
\dfrac{w_j(x)}{z-x}.
\end{equation}

Because of the form of $\Gamma_{I}$  we have that
\begin{equation*}
(T_i^*\,h)(\bs)=h(\bs+\be_i), \quad (T_i\,h)(\bs):=\begin{cases}
h(\bs-\be_i)\; \;\mbox{if $s_i\leq 1$}\\\\
0\;\; \mbox{if $s_i=0$},
\end{cases}
\end{equation*}
for functions
$h(\bs)\,(\bs\in\Gamma_I)$. It is clear that
\[
T_i^*\,T_i=\mathbb{I},\quad T_i\,T_i^*=(1-\delta_{s_i,
0})\,\mathbb{I},
\]
where $\mathbb{I}$ stands for the identity operator.
Sometimes it is helpful to think of the functions $h(\bs)$ as column vectors
$(h|_{s_i=0},h|_{s_i=1},h|_{s_i=2},\ldots)^T$. Thus, in this representation,
$T_i,\,T_i^*$ become the infinite-dimensional matrices
\[
T_i^*=\left(\begin{array}{ccccc}
0&1&0&\ldots&\ldots\\
0&0&1&0&\ldots\\
0&0&0&1&\ldots\\
\vdots&\vdots&\vdots&\vdots&\vdots
\end{array}\right)
,\quad T_i= \left(\begin{array}{ccccc}
0&0&0&\ldots&\ldots\\
1&0&0&0&\ldots\\
0&1&0&0&\ldots\\
\vdots&\vdots&\vdots&\vdots&\vdots
\end{array}\right)
.
\]

\subsection{The first system of string equations}

From the asymptotic expansion \eqref{mata} we have that as
$z\rightarrow\infty$
\begin{align*}
(T_i\,f)\,f^{-1}&=\Big(I+\dfrac{a(\bs-\be_i)}{z}+\mathcal{O}\Big(\dfrac{1}{z^2}\Big)\Big)\Big(z\,E_{0
}+\dfrac{E_{i}}{z}+I-E_{0}-E_{i}\Big)\Big(I-\dfrac{a(\bs)}{z}+\mathcal{O}\Big(\dfrac{1}{z^2}\Big)\Big)\\\\
&=z\,E_{0}+a(\bs-\be_i)\,E_{0}-E_{0}\,a(\bs)+I-E_{0}-E_{i}+\mathcal{O}\Big(\dfrac{1}{z}\Big),\quad \forall\bs\in\Gamma_{I}+\be_i,
\end{align*}
where we are denoting
\[
a(\bs):=a_1(\bs).
\]
Hence by applying Proposition 2 it follows that
\[
(T_i\,f)(\bs,z)=\Big(z\,E_{0}+a(\bs-\be_i)\,E_{0}-E_{0}\,a(\bs)+I-E_{0}-E_{i}\Big)\,f(\bs,z),\quad \bs\in\Gamma_{I}+\be_i
\]
which implies
\begin{equation}\label{sdos1}
(T_i E_{0}f)(\bs,z)=\Big((z-u_i(\bs))\,E_{0}-\sum_j\,
a_{0j}(\bs)\,E_{0j}\Big)f(\bs,z),\quad \forall\bs\in\Gamma_{I}+\be_i,
\end{equation}
where
\[
u_i(\bs):=a_{00}(\bs)-a_{00}(\bs-\be_i).
\]
Similarly one finds
\begin{equation}\label{sdos2}
(T_j^* E_{0}f)(\bs,z)=a_{0j}(\bs+\be_j)\,E_{0j}\,f(\bs,z),\quad
\forall \bs\in\Gamma_{I}.
\end{equation}
Note that as $\mbox{det}\,f(\bs,z)\equiv 1$ for all
$(\bs,z)\in \Gamma_{I}\times\C$ then , as a consequence of \eqref{sdos2} we
deduce that
\[
a_{0j}(\bs+\be_j)\neq 0,\quad \forall \bs\in\Gamma_{I}.
\]
If we now define
\begin{equation}\label{v1}
v_j(\bs):=\dfrac{a_{0j}(\bs)}{a_{0j}(\bs+\be_j)},\quad  \bs\in\Gamma_{I},
\end{equation}
then from \eqref{sdos1} it follows that \begin{pro}
The function $f$ satisfies the equations
\begin{equation}\label{string1uno}
z\,(E_{0}\,f)(\bs,z)=\Big(T_i+u_i(\bs)+\sum_{j}\,v_j(\bs)\,
T_j^*\Big)\,(E_{0}\,f)(\bs,z),
\end{equation}
for all $\bs\in\Gamma_{I}+\be_i$ and $i=1,\ldots, q$.
\end{pro}

As a consequence we get the following system of string equations

\begin{teh}
The multiple orthogonal polynomials of type I verify
\begin{equation}\label{string1pol1}
z\,\bA(\bn,z)=\Big(T_i+u_i(\bn)+\sum_{j}\,v_j(\bn)\,
T_j^*\Big)\,\bA(\bn,z),
\end{equation}
for all $\bn\in\Gamma_{I}+e_i$ and $i=1,\ldots,q$.
\end{teh}

For $q=1$ Eq.\eqref{string1pol1} reduces to the classical three-term recurrence relation for systems of orthogonal polynomials on the real line.

On the other hand Eq.\eqref{string1pol1} implies
\begin{equation}\label{re}
\bA(\bn-\be_i,z)-\bA(\bn-\be_j,z)=(a_{00}(\bn-\be_i)-a_{00}(\bn-\be_j))\,\bA(\bn,z),\quad \forall\bs\in\Gamma_I+\be_i+\be_j,\, i\neq j.
\end{equation}
The relations \eqref{string1pol1} and \eqref{re} lead  to a recursive method to construct the multiple orthogonal
polynomials of type I. Indeed, it is clear that for $\bn=n_i\be_i$ we have
$$
\bA(n_i\be_i,z)=A_1^{(i)}(n_i,z)\,\be_i=(0,\dots,0,A_1^{(i)}(n_i,z),0,\dots,0),
$$
where $A_1^{(i)}(n_i,z)$  are the orthogonal polynomials with respect to the weight  $w_i(x)$. Then starting from $A_1^{(i)}(n_i,z)$ and  using \eqref{re} we can generate the multiple orthogonal polynomials of type I for higher $q$.

\subsubsection*{Example}

 Let us denote by $I_{j,n}$ the moments with respect to the weight $w_j$
\begin{equation}\label{mom}
I_{j,n}:=\int_{\mathbb R} \frac{dx}{2\pi i}x^nw_j(x).
\end{equation}
We have that
$$A_1(1,z)=\frac{1}{I_{1,0}},\qquad A_1(2,z)=\frac{I_{1,0}z-I_{1,1}}{I_{1,0}I_{1,2}-I_{1,1}^2}.$$
The recurrence relation \eqref{string1pol1} for $q=1$ is
\begin{equation}\label{const1}
A_1(n+1,z)=\frac{a_{01}(n+1)}{a_{01}(n)}[(z+a_{00}(n-1)-a_{00}(n))A_1(n,z)-A_1(n-1,z)],\quad \forall n\geq2,
\end{equation}
where according to \eqref{sol1}
\begin{equation}\label{coef1}
a_{00}(n)=\int_{\mathbb R} \frac{dx}{2\pi i}x^nA(n,x)w_1(x).\end{equation}
Moreover, the normalization condition  gives us
\begin{equation}\label{coef2}
\frac{a_{01}(n)}{a_{01}(n+1)}=\int \frac{dx}{2\pi i}x^n[(x+a_{00}(n-1)-a_{00}(n))A(n,x)-A(n-1,x)]w_1(x).
\end{equation}
The system \eqref{const1}-\eqref{coef1}-\eqref{coef2} allows us
to construct the polynomials $A(n,z)$ for $n\geq3$. For example one gets
$$A_1(3,z)=\frac{\left(I_{1,1}^2-I_{1,0} I_{1,2}\right) z^2-I_{1,1}
   I_{1,3}+(I_{1,0} I_{1,3}-I_{1,1} I_{1,2}) z+I_{1,2}^2}{I_{1,2}^3-(2 I_{1,1}
   I_{1,3}+I_{1,0} I_{1,4}) I_{1,2}+I_{1,0} I_{1,3}^2+I_{1,1}^2 I_{1,4}}.$$

If we write \eqref{re}  in the form
\begin{equation}\label{const2}
\bA(\bn,z)=\frac{\bA(\bn-\be_i,z)-\bA(\bn-\be_j,z)}{a_{00}(\bn-\be_i)-a_{00}(\bn-\be_j)},\quad \bn\in\Gamma_I+\be_i+\be_j,
\end{equation}
and take into account that
\begin{equation}\label{coef3}
a_{00}(\bn)=\int_{\mathbb R}\frac{dx}{2\pi i}x^{|\bn|}\sum_{k=1}^qA_k(\bn,x)w_k(x).
\end{equation}
we can construct all the multiple orthogonal polynomials
of type I. Thus, for example for $q=2$ we obtain
\begin{align*}
\bA(1,1,z)&=\dfrac{1}{C_1}\,\Big(I_{2,0},-I_{1,0}\Big),\\  \\
\bA(2,1,z)&=\dfrac{1}{C_2}\,\Big(I_{1,2}I_{2,0} -I_{1,1}I_{2,1}+z(I_{1,0}I_{2,1}-I_{1,1}I_{2,0})\,,\,I_{1,1}^2-I_{1,0} I_{2,0}\Big),
\end{align*}
where
\begin{align*}
C_1:&=I_{1,1} I_{2,0}-I_{1,0}I_{2,1},\\\\
C_2:&=I_{2,2}I_{1,1}^2-I_{1,3}I_{2,0}I_{1,1}-I_{1,2}I_{2,1}I_{1,1}+I_{1,2}^2 I_{2,0}+I_{1,0}I_{1,3}I_{2,1}-I_{1,0}I_{1,2}I_{2,2}.
\end{align*}
\subsection{Lax operators}
%\subsubsection*{Example} Let us take $q=2$, $w_1(x)=e^{-\frac{a}{2}x^2+bx}$ and $w_2(x)=e^{-\frac{c}{2}x^2+dx}$. Then we have
%$$\everymath{\displaystyle}\begin{array}{lll}
%\bA(1,1,z)&=&\left(\frac{i a^{3/2} c e^{-\frac{b^2}{2 a}} \sqrt{2 \pi }}{b c-a d},
%              \frac{i a c^{3/2} e^{-\frac{d^2}{2 c}} \sqrt{2 \pi }}{a d-b c}\right),\\  \\
%\bA(2,1,z)&=&\left(-\frac{i a^{3/2} c e^{-\frac{b^2}{2 a}} \sqrt{2 \pi }
%              \left(d z a^2+(-b z c+c-b d) a+b^2c\right)}{\left(d^2+c\right) a^2-c (c+2 b d) a+b^2c^2},
%              \frac{i a^2 c^{5/2} e^{-\frac{d^2}{2 c}} \sqrt{2 \pi}}{\left(d^2+c\right) a^2-c (c+2 b d) a+b^2 c^2}\right).

%\end{array}$$

\vspace{0.3cm}
The functions $f_{0i}(\bs,z)=A_i(\bs,z)$ can be written as series expansions of the form
$$
f_{0i}(\bs,z)=\left(\frac{\alpha_{i1}(\bs)}{z}+\frac{\alpha_{i2}(\bs)}{z^2}+\cdots\right)\,z^{s_i},
$$
where
\begin{equation}\label{alp}
\alpha_{in}(\bs)=0,\,\forall n\geq{s_i+1}.
\end{equation}
On the other hand, it is easy to see that
\begin{equation}\label{tnes}
T_i^n\,z^{s_i}=\dfrac{1}{z^n}\,(z^{s_i}-\sum_{k=0}^{n-1}z^k\,
\delta_{s_i-k,0}),
\end{equation}
Hence, from \eqref{alp} and \eqref{tnes} it is clear that
\[
\dfrac{\alpha_{i,n+1}(\bs+\be_i)}{z^n}\,z^{s_i}=\alpha_{i,n+1}(\bs+\be_i)
T_i^n\,z^{s_i},\quad \forall n\geq 1,
\]
so that we may write
\[
f_{0i}(\bs+\be_i,z)=(G_{i}\,\xi_{i})(\bs,z),\quad \xi_{i}(\bs,z):=z^{s_i},\quad
\bs\in\Gamma_{I},
\]
where the symbols $G_i$ are dressing operators defined by the expansions
\begin{equation}\label{dreuno}
G_i=\sum_{n\geq 1}\alpha_{in}(\bs+\be_i)\,T_i^{n-1},\quad
\alpha_{in}(\bs+\be_i):=(a_{n})_{0i}(\bs+\be_i),
\end{equation}
or, equivalently, by the triangular matrices
\[
G_i=\left(\begin{array}{ccccc}
G_{00}&0&0&\ldots&\ldots\\
G_{10}&G_{11}&0&0&\ldots\\
G_{20}&G_{21}&G_{22}&0&\ldots\\
\vdots&\vdots&\vdots&\vdots&\vdots
\end{array}\right),\quad G_{nm}=\alpha_{i,n-m+1}(\bs+\be_i)\Big|_
{s_i=m}.
\]
The inverse operators can be written as
\[
G_i^{-1}:=\sum_{n\geq 1}\beta_{in}(\bs)\,T_i^{n-1},\quad
\beta_{i1}(\bs)=\dfrac{1}{\alpha_{i1}(\bs+\be_i)}=\dfrac{1}{a_{0i}
(\bs+\be_i)}.
\]

We  define the Lax operators $\bZ_i$ by
\begin{equation}\label{lax2}
\bZ_i:=G_i\,T_i^*\,G_i^{-1}.
\end{equation}
It follows at once that they can be expanded as
\begin{equation}\label{lax2euno}
\bZ_i=\gamma_i(\bs)\,T_i^*+\sum_{n\geq 0}\gamma_{in}(\bs)\,T_i^n,
\end{equation}
where
\begin{equation}
\gamma_i(\bs)=\alpha_{i1}(\bs+\be_i)\,(T_i^*\,\beta_{i1})(\bs)=
\dfrac{\alpha_{i1}(\bs+\be_i)}{\alpha_{i1}(\bs+2\,\be_i)}= v_i(\bs+\be_i).
\end{equation}
\begin{pro}
The functions $f_{0i}$ satisfy the equations
\begin{equation}\label{ste11uno}
z\,f_{0i}(\bs+\be_i,z)=(\bZ_i\,f_{0i})(\bs+\be_i,z),\quad \forall
\bs\in\Gamma_{I}.
\end{equation}
\end{pro}
\begin{proof}
%\begin{equation}\label{ste11uno}
%z\,f_{0i}(\bs+\be_i,z)=(\bZ_i\,f_{0i})(\bs+\be_i,z)
%=\Big(T_{j}+u_{j}(\bs+\be_i)+\sum_{k}\,v_k(\bs+\be_i)\, T_k^*\Big)\,f_{0i}(\bs+\be_i,z),\end{equation}
%for all $\bs\in\Gamma_{I}+\sum_{k\neq i}\be_k$ and  $i,j=1,\ldots,q$.
%\end{pro}
From the definition of $G_i$ we have
\[
z\,f_{0i}(\bs+\be_i,z)=G_i\,(z\,\xi_i)=(G_i\,T_i^*)\,(\xi_i) =
(\bZ_i\,f_{0i})(\bs+\be_i,z).
\]
%The rest of the proof follows from the system of string equations
%\eqref{string1uno}.
\end{proof}

%\begin{align*}
%f_{00}(\bs,z)&=(G_0^{(i)}\,\xi_{0})(\bs,z),\quad G_0^{(i)}=1+\sum_{n\geq 1}\alpha_{0n}(\bs)\,(T_i^*)^{n},\\
%\xi_{0}(\bs,z):&=z^{s_0},\quad \alpha_{0n}(\bs):=(a_{n})_{00}(\bs),\quad \bs\in\Gamma_{I}.
%\end{align*}
%The corresponding Lax operators are defined by
%\begin{equation}\label{zec1}
%\bZ_0^{(i)}:=G_0^{(i)}\,T_i\,(G_0^{(i)})^{-1}=T_{i}+u_{i}(\bs)+\sum_{n\geq 1}\gamma_{0n}^{(i)}(\bs)\, (T_i^*)^n.,
%\end{equation}
%and taking into account that
%\[
%z^{s_0+1}=T_i z^{s_0}+\delta_{s_i, 0}\,z^{s_0+1}; \quad (T_i^*)^n\,\delta_{s_i 0}=0,\quad \forall n\geq
%1,\;\bs\in\Gamma_{II},
%\]
%we have that
%\[
%z\,f_{00}(\bs,z)=(\bZ_0^{(i)}\,f_{00})(\bs,z)+\delta_{s_i 0}\,z^{s_0+1},\quad \forall \bs\in\Gamma_{I},
%\]
%and therefore we deduce that
%\begin{pro} The function $f_{00}$ satisfies the equations
%\begin{equation}\label{ste1cero}
%z\,f_{00}(\bs,z)=(\bZ_0^{(i)}\,f_{00})(\bs,z),\quad \forall \bs\in\Gamma_{I}+\be_i,\;i=1,\ldots,q.
%\end{equation}
%\end{pro}

%Thus we will say that the Lax operators $\bZ_i$ satisfy the system
%of weak string equations
%\begin{equation}\label{first2uno}
%\mathcal{Z}_{0}\approx \mathcal{Z}_{1}\approx\ldots\approx
%\mathcal{Z}_{q}.
%\end{equation}

\subsection{The second system of string equations}
Let us consider diagonal solutions
\[
\Phi(z)=\sum_{\alpha} \phi_{\alpha}(z)\,E_{\alpha}
\]
of the condition \eqref{inv} corresponding to the
function $g(z)$ of \eqref{gtypeI}. They are characterized by
\[
\partial_z\,w_i-\phi_0\,w_i+\phi_i\,w_i=0,\quad i=1,\ldots,q.
\]
In this way, by setting $\phi_0\equiv 0$ we get
\[
\Phi(z)=\sum_{i} V'(\bc_i,z)\,E_{i}.
\]
The corresponding covariant derivative is
\begin{equation}\label{cov1uno}
D_z\,f:=\partial_z f-\sum_i
V'(\bc_i,z)\,f\,E_{i}.
\end{equation}
Hence we have
\begin{equation}\label{cov2uno}
D_z(\,E_{0}\,f)=\partial_z f_{00}\,E_{0}+\sum_i (\partial_z
f_{0i}-V'(\bc_i,z)\,f_{0i})\,E_{0i}.
\end{equation}

It is clear that \eqref{ste11uno} implies
\begin{equation}\label{powuno}
z^n\,f_{0i}(\bs+\be_i,z)=
(\bZ_i^n\,f_{0i})(\bs+\be_i,z),\quad \forall \bs\in\Gamma_{I}.
\end{equation}
On the other hand, as $z\rightarrow\infty$
\begin{align}\label{masuno}
\nonumber &((T_j^*)^n\,f_{0\alpha})(\bs,z)=\begin{cases}
\mathcal{O}\Big(\dfrac{1}{z^n}\Big)\,z^{s_0},\,
\mbox{for $\alpha=0$},
\\\\
\mathcal{O}\Big(\dfrac{1}{z}\Big)\,z^{s_i},\,
\mbox{for $\alpha=i\neq j$ },
\end{cases},\quad  n\geq 1,\\\\
\nonumber &(T_i^n\,f_{0i})(\bs,z)=\begin{cases}
\mathcal{O}\Big(\dfrac{1}{z^{n+1}}\Big)\,z^{s_i},\,
\mbox{for $s_i\geq n$},
\\\\
0,\, \mbox{for $s_i<n$}
\end{cases},\quad  n\geq 0.
\end{align}
\begin{pro}
The function $f$ satisfies the equation
\begin{equation}\label{string2uno}
(D_z+{\cal H})\,(E_{0}\,f)(\bs,z)=0,
\quad \forall \bs\in\Gamma_{I}+\sum_j \be_j,
\end{equation}
where  ${\cal H}$ is the operator
\begin{equation}\label{oper}
\mathcal{H}:=\sum_{j
=1}^q V'(\bc_j,\mathcal{Z}_j)_{(j,+)}
\end{equation}
\end{pro}
\begin{proof}
Given $\bs\in\Gamma_{I}+\sum_j \be_j$ let us denote
\[
\bs^{(i)}:=\bs-\be_i\in\Gamma_{I}+\sum_{k\neq i}\be_k.
\]
From \eqref{cov2uno} it follows that
\begin{equation}\label{covf22uno}\everymath{\displaystyle}
\begin{array}{l}
(D_z+{\cal H})\,(E_{0}\,f)\,=
\left[\partial_zf_{00}+\sum_{j=1}^qV'(\bc_j,\mathcal{Z}_j)
_{(j,+)}\,f_{00}\right]\,E_{0}\\  \\
\qquad\qquad +\sum_{i=1}^q\left[\partial_zf_{0i}+ \sum_{j=1}^q V'(\bc_j,\mathcal{Z}_j)_{(j,+)}\,f_{0i}
 -V'(\bc_i,z)\,f_{0i}\right]E_{0i}
\end{array}\end{equation}
Now from  \eqref{masuno} we have
\begin{align*}
&\partial_zf_{00}+\sum_{j=1}^qV'(\bc_j,\mathcal{Z}_j)
_{(j,+)}\,f_{00}=\mathcal{O}\Big(\dfrac{1}{z}\Big)\,z^{s_0},\\
&\partial_zf_{0i}+ \sum_{j\neq i} V'(\bc_j,\mathcal{Z}_j)_{(j,+)}\,f_{0i}=\mathcal{O}\Big(\dfrac{1}{z}\Big)\,z^{s_i}.
\\
&\Big(V'(\bc_i,\mathcal{Z}_i)_{(i,+)}-V'(\bc_i,z)\Big)\,f_{0i}(\bs,z)=
\Big(V'(\bc_i,\mathcal{Z}_i)-V'(\bc_i,z)\Big)\,f_{0i}(\bs,z)+\mathcal{O}\Big(\dfrac{1}{z}\Big)\,z^{s_i},
\end{align*}
Moreover, from \eqref{powuno} it is clear that
\begin{align*}
&\Big(V'(\bc_i,\mathcal{Z}_i)-V'(\bc_i,z)\Big)\,f_{0i}(\bs,z)=
\Big(V'(\bc_i,\mathcal{Z}_i)-V'(\bc_i,z)\Big)\,f_{0i}(\bs^{(i)}+\be_i,z)\\\\
&
=\sum_{n\geq 1}n\,c_{in}\,\Big(\mathcal{Z}_i^{n-1}-z^{n-1}\Big)\,f_{0i}(\bs^{(i)}+\be_i,z)=0
\end{align*}
Therefore we find 
\[
(D_z+{\cal H})\,(E_{0}\,f)(\bs,z)=
{\cal
O}(\frac{1}{z})\,f_0(\bs,z),\quad z\rightarrow\infty.
\]
The first member $\tilde{f}:=(D_z+{\cal H})\,(E_{0}\,f)$
of this equation is a solution of \eqref{rhp} for all $\bs\in\Gamma_{I}+\sum_j\be_j$ and $\tilde{f}(\bs,z)\,f(\bs,z)^{-1}\rightarrow 0$ as $z\rightarrow
\infty$. Therefore, the statement of Proposition 2 implies $\tilde{f}\equiv 0$
.
\end{proof}

As a consequence we deduce the following system of string equations
\begin{teh}
The multiple orthogonal polynomials of type I verify
\begin{equation}\label{string1poluno}
\partial_z\,A_i(\bn,z)=V'(\bc_i,\bZ_i)\,A_i(\bn,z)-\sum_{j
=1}^q V'(\bc_j,\mathcal{Z}_j)_{(j,+)}\,A_i(\bn,z),
\end{equation}
for all $\bn\in\Gamma_I+\sum_k \be_k$ and $i=1,\ldots,q$.
\end{teh}

\subsection{Orlov operators}

We define  the Orlov operators $\bM_i$ by
\begin{equation}\label{or2uno}
\bM_i:=G_i\cdot s_i\cdot T_i\cdot G_i^{-1}.
\end{equation}
They satisfy $[\bZ_i,\bM_i]=\mathbb{I}$ and can be expanded as
\begin{equation}\label{or2e}
\bM_i=\sum_{n\geq 1}\mu_{in}(\bs)\,T_i^n,
\end{equation}
where
\begin{equation}
\mu_{i1}(\bs)=\dfrac{s_i}{v_i(\bs)}.
\end{equation}
\begin{pro}
The functions $f_{0i}$ satisfy the equations
\begin{equation}\label{ste2uno}
\partial_z\,f_{0i}(\bs+\be_i,z)=(\bM_i\,f_{0i})(\bs,z+\be_i),\quad
\forall \bs\in\Gamma_{I}.
\end{equation}
\end{pro}
%=\Big(-\mathcal{H}+V'(\bc_i,\bZ_i)\Big)\,f_{0i}(\bs,z),

\begin{proof}
From the definition of $G_i$ we have
\[
\partial_z\,f_{0i}(\bs+\be_i,z)=G_i\,(s_i\,z^{-1}\,\xi_i)=G\cdot s_i (T_i\,\xi_i) =
G_i\cdot s_i\cdot T_i\cdot G_i^{-1} f_{0i}=\bM_i\,f_{0i}(\bs+\be_i,z).
\]

\end{proof}

%If we define
%\begin{equation}\label{s2uno}
%\bM_0:=\mathcal{H}=\sum_{j
%=1}^q V'(\bc_j,\mathcal{Z}_j)_{(j,+)},
%\end{equation}

%  we get
%\begin{equation}\label{s2uno}
%\bM_0\,f_{00}=\mathcal{H}\,f_{00},\quad \bM_i\,f_{0i}=\Big(\mathcal{H}-V'(\bc_i,\bZ_i)\Big)\,f_{0i},
%\end{equation}
%for all $\bs\in\Gamma_{I}+\sum_j\be_j$.
%Thus we will say that the Orlov operators  satisfy a system
%of weak string equations
%\begin{equation}\label{first2dos}
%\bM_{0}\approx \bM_{1}+V'(\bc_1,\bZ_1)\approx\ldots\approx
%\bM_{q}+V'(\bc_q,\bZ_q).
%\end{equation}

%\subsubsection*{Example} Let us take $q=2$ and
%$$\bc_1=(-\frac{a}{2},b,0,0,\dots),\quad \bc_2=(-\frac{c}{2},d,0,0,\dots).$$
%Then \eqref{string2typeIpol} reads
%$$\everymath{\displaystyle}\begin{array}{lll}
%\partial_zA_1(\bs,z)+(-az+b)A_1(\bs,z)&=&-a\frac{a_{01}(\bs)}{a_{01}(\bs+\be_1)}A_1(\bs+\be_1,z)
%-c\frac{a_{02}(\bs)}{a_{02}(\bs+\be_2)}A_1(\bs+\be_2,z),\\  \\
%\partial_zA_2(\bs,z)+(-cz+d)A_2(\bs,z)&=&-a\frac{a_{01}(\bs)}{a_{01}(\bs+\be_1)}A_2(\bs+\be_1,z)
%-c\frac{a_{02}(\bs)}{a_{02}(\bs+\be_2)}A_2(\bs+\be_2,z),
%\end{array}$$
%where $a_{01}(\bs)=\mbox{coeff}[A_1(\bs,z),z^{s_1-1}]$ and  $a_{02}(\bs)=\mbox{coeff}[A_2(\bs,z),z^{s_2-1}]$.

\section{Multiple orthogonal polynomials of type II}

We consider now $q$ exponential weights $w_i$ on the real line \[
w_i(x):=e^{V(\bc_i,x)},\quad
\bc_i=(c_{i1},c_{i2},\ldots)\in\C^{\infty}.
\]
Note the difference in the sign of the exponents with respect to the weights for multiple orthogonal polynomials of type I.
Given $\bn=(n_1,\ldots,n_q)\in\N^q$, the associated
type II monic orthogonal polynomial $P(\bn,x)=x^{|\bn|}+\cdots$  is
determined by the conditions
\begin{equation*}
\int_{\mathbb{R}} P(\bn,x)\,w_i(x)\,x^j\,\d x=0,\quad
j=0,\ldots,n_i-1.
\end{equation*}
We assume that all the multi-indices $\bn$ are strongly normal
\cite{kuij} so that $P(\bn,z)$ is unique.

The  RH problem  for the multiple orthogonal polynomials of type II is determined by
\cite{kuij}
\begin{equation}\label{fokg1}
g(z)=\left(
\begin{array}{ccccc}
1&w_1(z)&w_2(z)&\ldots&w_q(z)\\
0&1&0&\ldots&0\\
\vdots&\vdots&\vdots&\vdots&\vdots\\
0&0&\ldots&0&1
\end{array}
\right),
\end{equation}
Its fundamental solution $f(\bs,z)$ exists on the domain
\begin{equation}\label{d2}
\Gamma_{II}=\{\bs\in\Z^{q}\,:\, s_i\leq 0,\,\forall i=1,\ldots,q\},
\end{equation}
For $ s_i\leq -1 \,\forall i$, it is given by
\begin{align}\label{sol2}
\nonumber &f(\bs,z)=\left(
\begin{array}{cc}
P(\bn,z)&\bR(\bn,z)\\
d_1\,P(\bn-\be_1,z)&d_1\,\bR(\bn-\be_1,z)\\
\vdots&\vdots\\
d_q\,P(\bn-\be_q,z)&d_q\,\bR(\bn-\be_q,z)
\end{array}
\right),\quad \bs=-\bn,\\\\
\nonumber & R_j(\bn,z):=\int_{\mathbb{R}}\dfrac{\d x}{2\pi i}
\dfrac{P(\bn,x)\,w_j(x)}{x-z}, \quad\quad\quad
\dfrac{1}{d_j}:=-\int_{\mathbb{R}}\dfrac{\d x}{2\pi i}
P(\bn-\be_j,x)\,w_j(x)\,x^{n_j-1}.
\end{align}
For the remaining cases, in which  one or several $s_i$ vanish,
one must insert the following corresponding row substitutions in
\eqref{sol2}
\begin{equation}\label{sol2a}
(d_i\,P(\bn-\be_i,z)\quad d_i\,\bR(\bn-\be_i,z))\;\longrightarrow\;
(0\quad \be_i).
\end{equation}
In particular
\begin{equation*}
f(\boldsymbol{0},z)=\left(
\begin{array}{ccccc}
1&R_1(z)&R_2(z)&\cdots&R_q(z)\\
0&1&0&\cdots&0\\
\vdots&\vdots&\vdots&\vdots\\
0&0&0&\cdots&1
\end{array}
\right),\quad R_j(z):=\int_{\mathbb{R}}\dfrac{\d x}{2\pi i}
\dfrac{w_j(x)}{x-z}.
\end{equation*}

In view of \eqref{d2}  we have that
\begin{equation*}
(T_i\,h)(\bs)=h(\bs-\be_i), \quad (T_i^*\,h)(\bs):=\begin{cases}
h(\bs+\be_i)\; \;\mbox{if $s_i\leq -1$}\\\\
0\;\; \mbox{if $s_i=0$},
\end{cases}
\end{equation*}
for functions
$h(\bs)\,(\bs\in\Gamma_{II})$. Note also that
\[
T_i\,T_i^*=\mathbb{I},\quad T_i^*\,T_i=(1-\delta_{s_i,
0})\,\mathbb{I}
\]
where $\mathbb{I}$ stands for the identity operator. If we think of $h(\bs)$ as a column vector
$(h|_{s_i=0},h|_{s_i=-1},h|_{s_i=-2},\ldots)^T$, then
$T_i,\,T_i^*$ are represented by the infinite-dimensional matrices
\[
T_i=\left(\begin{array}{ccccc}
0&1&0&\ldots&\ldots\\
0&0&1&0&\ldots\\
0&0&0&1&\ldots\\
\vdots&\vdots&\vdots&\vdots&\vdots
\end{array}\right)
,\quad T_i^*= \left(\begin{array}{ccccc}
0&0&0&\ldots&\ldots\\
1&0&0&0&\ldots\\
0&1&0&0&\ldots\\
\vdots&\vdots&\vdots&\vdots&\vdots
\end{array}\right)
.
\]

\subsection{The first system of string equations}

The same analysis as in the  subsection 3.1 leads now to the equations
\begin{equation}\label{s11}
(T_i E_{0}f)(\bs,z)=\Big((z-u_i(\bs))\,E_{0}-\sum_j\,
a_{0j}(\bs)\,E_{0j}\Big)f(\bs,z),\quad \forall \bs\in\Gamma_{II},
\end{equation}
where
\[
u_i(\bs):=a_{00}(\bs)-a_{00}(\bs-\be_i).
\]
Similarly one finds
\begin{equation}\label{s12}
(T_j^* E_{0}f)(\bs,z)=a_{0j}(\bs+\be_j)\,E_{0j}\,f(\bs,z),\quad
\forall \bs\in\Gamma_{II}-\be_i,
\end{equation}
and taking into account that  $\mbox{det}\,f(\bs,z)\equiv 1$ for all
$(\bs,z)\in \Gamma_{II}\times\C$, from \eqref{s12} we obtain
\[
a_{0j}(\bs)\neq 0,\quad \forall \bs\in\Gamma_{II}.
\]
Now we define
\begin{equation}
v_j(\bs):=\begin{cases}\dfrac{a_{0j}(\bs)}{a_{0j}(\bs+\be_j)},\quad \bs\in\Gamma_{II}-\be_i,\\\\
0,\quad \mbox{for $s_j=0$}.
\end{cases}
\end{equation}
Notice that the functions $v_j(\bs)\, (\bs\in\Gamma_{I})$  for multiple orthogonal polynomials of type I defined in \eqref{v1} also satisfies $v_j(\bs)=0$ for $s_j=0$.

If we now
recall that according to \eqref{sol2a}
\[
E_{0j}\,f(\bs,z)=E_{0j},\quad \mbox{for $s_j=0$},
\]
from \eqref{s11} it follows that
\begin{pro}
The function $f$ satisfies the equations
\begin{equation}\label{string1}
z\,E_{0}\,f(\bs,z)=\Big(T_i+u_i(\bs)+\sum_{j}\,v_j(\bs)\,
T_j^*\Big)\,(E_{0}\,f)(\bs,z)+\sum_j
\delta_{s_j,0}\,a_{0j}(\bs)\,E_{0j},
\end{equation}
for all $\bs\in\Gamma_{II}$ and $i=1\ldots q$.
\end{pro}
As a consequence we get the string equations
\begin{teh}
The multiple orthogonal polynomials of type II verify
\begin{equation}\label{string1poldos}
 z\,P(\bn,z)=\Big(T_i+u_i(-\bn)+\sum_{j}\,v_j(-\bn)\,
T_j^*\Big)\,P(\bn,z),
\end{equation}
for all $\bn$ and $i=1,\ldots, q$.

\end{teh}

These equation provide a recursive method to construct multiple orthogonal polynomials
of type II. We may write \eqref{string1poldos} as
\begin{equation}\label{constr2}
P(\bn+\be_j,z)-a_{00}(\bn+\be_j)P(\bn,z)=(z-a_{00}(\bn))P(\bn,z)
-\sum_{k=1,n_k\geq1}^q\frac{a_{0k}(\bn)}{a_{0k}(\bn-e_k)}P(\bn-e_k,z),
\end{equation}
where, according to\eqref{sol2}, we have that
\begin{equation}\label{coef21}\everymath{\displaystyle}\begin{array}{lll}
a_{00}(\bn)&=&\mbox{coeff}[P(\bn,z),z^{|\bn|-1}],\\  \\
a_{0k}(\bn)&=&-\int_{\mathbb R}\frac{dx}{2\pi i}P(\bn,x)x^{n_k}w_k(x)dx.
\end{array}\end{equation}
On the other hand,  multiplying the equation \eqref{constr2} by $z^{n_j}w_j(z)$, integrating on ${\mathbb R}$ and
using the orthogonality condition for $P(\bn+\be_j,z)$, we obtain
$$\everymath{\displaystyle}\begin{array}{l}
a_{00}(\bn+e_j)\left[-\int_{\mathbb R}\frac{dx}{2\pi i}P(\bn,x)x^{n_j}w_j(x)\right]=\\  \\
\qquad\qquad\qquad\int_{\mathbb R}\left[(x-a_{00}(\bn))P(\bn,x)-
\sum_{k=1,n_k\geq1}^q\frac{a_{0k}(\bn)}{a_{0k}(\bn-e_k)}P(\bn-e_k,x)\right]x^{n_j}w_j(x)dx,
\end{array}$$
so that
\begin{equation}\label{coef22}
a_{00}(\bn+e_j)=\frac{1}{a_{0j}(\bn)}\int_{\mathbb R}\left[(x-a_{00}(\bn))P(\bn,x)-
\sum_{k=1,n_k\geq1}^q\frac{a_{0k}(\bn)}{a_{0k}(\bn-e_k)}P(\bn-e_k,x)\right]x^{n_j}w_j(x)dx.
\end{equation}
The system \eqref{constr2}-\eqref{coef21}-\eqref{coef22} determines  the multiple
orthogonal polynomials of type II in terms of the moments $I_{j,n}$ .

\subsubsection*{Example}

For $q=1$ is clear that
$$P(0,z)=1,\qquad P(1,z)=z-\frac{I_{1,1}}{I_{1,0}}.$$
From \eqref{constr2}-\eqref{coef21}-\eqref{coef22} we easily obtain that
$$\everymath{\displaystyle}\begin{array}{lll}
P(2,z)&=&z^2+\frac{(I_{1,0}I_{1,3}-I_{1,1}I_{1,2})z}{I_{1,1}^2-I_{1,0}I_{1,2}}+
\frac{I_{1,2}^2-I_{1,1}I_{1,3}}{I_{1,1}^2-I_{1,0}I_{1,2}},\\  \\
P(3,z)&=&z^3+\frac{\left(-I_{1,5} I_{1,1}^2+I_{1,3}^2
I_{1,1}+I_{1,2} I_{1,4}
   I_{1,1}-I_{1,2}^2 I_{1,3}-I_{1,0} I_{1,3} I_{1,4}+I_{1,0} I_{1,2}
   I_{1,5}\right) z^2}{I_{1,2}^3-(2 I_{1,1} I_{1,3}+I_{1,0} I_{1,4})
   I_{1,2}+I_{1,0} I_{1,3}^2+I_{1,1}^2 I_{1,4}}\\  \\
   &  &+\frac{\left(-I_{1,4}
   I_{1,2}^2+I_{1,3}^2 I_{1,2}+I_{1,1} I_{1,5} I_{1,2}+I_{1,0}
   I_{1,4}^2-I_{1,1} I_{1,3} I_{1,4}-I_{1,0} I_{1,3} I_{1,5}\right)
   z}{I_{1,2}^3-2 I_{1,1} I_{1,3} I_{1,2}-I_{1,0} I_{1,4}
   I_{1,2}+I_{1,0} I_{1,3}^2+I_{1,1}^2 I_{1,4}}\\  \\
   &  &-\frac{I_{1,3}^3-2 I_{1,2}
   I_{1,4} I_{1,3}-I_{1,1} I_{1,5} I_{1,3}+I_{1,1} I_{1,4}^2+I_{1,2}^2
   I_{1,5}}{I_{1,2}^3-2 I_{1,1} I_{1,3} I_{1,2}-I_{1,0} I_{1,4}
   I_{1,2}+I_{1,0} I_{1,3}^2+I_{1,1}^2 I_{1,4}}.
\end{array}$$
To determine the orthogonal polynomials for $q\geq2$ we use the property
$$P(n_i\be_i,z)=P^{(i)}(n_i,z),$$
where $P^{(i)}(n_i,z)$ are the orthogonal polynomials for $q=1$ with respect to the weight $w_i(x)$. For example for $q=2$ and $j=2$, Eq.\eqref{constr2} yields

$$\everymath{\displaystyle}\begin{array}{lll}
P(1,1,z)&=&z^2+\frac{(I_{1,2} I_{2,0}-I_{1,0} I_{2,2}) z}{I_{1,0}
I_{2,1}-I_{1,1}
   I_{2,0}}+\frac{I_{1,2} I_{2,1}-I_{1,1} I_{2,2}}{I_{1,1}
   I_{2,0}-I_{1,0} I_{2,1}},\\  \\
P(2,1,z)&=&z^3+\frac{\left(-I_{2,3} I_{1,1}^2+I_{1,4} I_{2,0}
I_{1,1}+I_{1,3} I_{2,1}
   I_{1,1}-I_{1,2} I_{1,3} I_{2,0}-I_{1,0} I_{1,4} I_{2,1}+I_{1,0}
   I_{1,2} I_{2,3}\right) z^2}{I_{2,2} I_{1,1}^2-I_{1,3} I_{2,0}
   I_{1,1}+I_{1,2}^2 I_{2,0}+I_{1,0} I_{1,3} I_{2,1}-I_{1,2} (I_{1,1}
   I_{2,1}+I_{1,0} I_{2,2})}\\  \\
   &  &+\frac{\left(I_{2,0} I_{1,3}^2-I_{1,1} I_{2,2}
   I_{1,3}-I_{1,0} I_{2,3} I_{1,3}-I_{1,2} I_{1,4} I_{2,0}+I_{1,0}
   I_{1,4} I_{2,2}+I_{1,1} I_{1,2} I_{2,3}\right) z}{I_{2,2}
   I_{1,1}^2-I_{1,3} I_{2,0} I_{1,1}-I_{1,2} I_{2,1} I_{1,1}+I_{1,2}^2
   I_{2,0}+I_{1,0} I_{1,3} I_{2,1}-I_{1,0} I_{1,2}
   I_{2,2}}\\  \\
   & &+\frac{I_{2,3} I_{1,2}^2-I_{1,4} I_{2,1} I_{1,2}+I_{1,3}^2
   I_{2,1}+I_{1,1} I_{1,4} I_{2,2}-I_{1,3} (I_{1,2} I_{2,2}+I_{1,1}
   I_{2,3})}{-I_{2,2} I_{1,1}^2+I_{1,3} I_{2,0} I_{1,1}+I_{1,2} I_{2,1}
   I_{1,1}-I_{1,2}^2 I_{2,0}-I_{1,0} I_{1,3} I_{2,1}+I_{1,0} I_{1,2}
   I_{2,2}}.
\end{array}$$
%Proceeding in the same way for $q=3$ we get
%$$\everymath{\displaystyle}\begin{array}{lll}
%P(1,1,1,z)&=&z^3+\frac{(I_{1,3} I_{2,1} I_{3,0}-I_{1,1}I_{2,3}I_{3,0}-I_{1,3}
%   I_{2,0} I_{3,1}+I_{1,0} I_{2,3} I_{3,1}+I_{1,1} I_{2,0}I_{3,3}-I_{1,0} I_{2,1} I_{3,3}) z^2}{-I_{1,2} I_{2,1}
%   I_{3,0}+I_{1,1} I_{2,2} I_{3,0}+I_{1,2} I_{2,0} I_{3,1}-I_{1,0}
%   I_{2,2} I_{3,1}-I_{1,1} I_{2,0} I_{3,2}+I_{1,0} I_{2,1}
%   I_{3,2}}\\  \\
%   & &+\frac{(I_{1,3} I_{2,2} I_{3,0}-I_{1,2} I_{2,3}
%   I_{3,0}-I_{1,3} I_{2,0} I_{3,2}+I_{1,0} I_{2,3} I_{3,2}+I_{1,2}
%   I_{2,0} I_{3,3}-I_{1,0} I_{2,2} I_{3,3}) z}{I_{1,2} I_{2,1}
%   I_{3,0}-I_{1,1} I_{2,2} I_{3,0}-I_{1,2} I_{2,0} I_{3,1}+I_{1,0}
%   I_{2,2} I_{3,1}+I_{1,1} I_{2,0} I_{3,2}-I_{1,0} I_{2,1}
%   I_{3,2}}\\  \\
%   & &+\frac{I_{1,3} I_{2,2} I_{3,1}-I_{1,2} I_{2,3}
%   I_{3,1}-I_{1,3} I_{2,1} I_{3,2}+I_{1,1} I_{2,3} I_{3,2}+I_{1,2}
%   I_{2,1} I_{3,3}-I_{1,1} I_{2,2} I_{3,3}}{-I_{1,2} I_{2,1}
%   I_{3,0}+I_{1,1} I_{2,2} I_{3,0}+I_{1,2} I_{2,0} I_{3,1}-I_{1,0}
%   I_{2,2} I_{3,1}-I_{1,1} I_{2,0} I_{3,2}+I_{1,0} I_{2,1}
%   I_{3,2}}.
%\end{array}$$

\subsection{Lax operators}

Let us   introduce
dressing operators $G_{i}$ according to
\[
f_{0i}(\bs,z)=(G_{i}\,\xi_{i})(\bs,z),\quad G_i:=\sum_{n\geq 0}\alpha_{in}(\bs)\,T_i^n,\quad
\alpha_{in}(\bs):=(a_{n+1})_{0i}(\bs),
\]
where $\bs\in\Gamma_{II}$ and $\xi_{i}(\bs,z):=z^{s_i-1}$. In the matrix representation they are given by the triangular matrices

%\begin{equation}\label{dre}
%G_i:=\sum_{n\geq 0}\alpha_{in}(\bs)\,T_i^n,\quad
%\alpha_{in}(\bs):=(a_{n+1})_{0i}(\bs),
%\end{equation}
\[
G_i=\left(\begin{array}{ccccc}
G_{00}&G_{01}&G_{02}&\ldots&\ldots\\
0&G_{11}&G_{12}&G_{13}&\ldots\\
0&0&G_{22}&G_{23}&\ldots\\
\vdots&\vdots&\vdots&\vdots&\vdots
\end{array}\right),\quad G_{nm}=\alpha_{i,m-n}(\bs)\Big|_{s_i=-m}.
\]
The corresponding inverse operators are characterized by  expansions of the form
\[
G_i^{-1}:=\sum_{n\geq 0}\beta_{in}(\bs)\,T_i^n,\quad
\beta_{i0}(\bs)=\dfrac{1}{\alpha_{i0}(\bs)}=\dfrac{1}{a_{0i}(\bs)}.
\]

We define the Lax operators $\bZ_i$ by
\begin{equation}\label{lax2}
\bZ_i:=G_i\,T_i^*\,G_i^{-1}.
\end{equation}
It follows at once that they can be expanded as
\begin{equation}\label{lax2e}
\bZ_i=\gamma_i(\bs)\,T_i^*+\sum_{n\geq 0}\gamma_{in}(\bs)\,T_i^n,
\end{equation}
where
\begin{equation}
\gamma_i(\bs)=\alpha_{i0}(\bs)\,(T_i^*\beta_{i0})(\bs)= v_i(\bs).
\end{equation}
\begin{pro}
The functions $f_{0i}$ satisfy the equations
\begin{equation}\label{ste11}
z\,f_{0i}(\bs,z)=(\bZ_i\,f_{0i})(\bs,z)+a_{0i}(\bs)\,\delta_{s_i 0},\quad \forall \bs\in\Gamma_{II}.
\end{equation}
\end{pro}

\begin{proof}
From the definition of $G_i$ we have
\[
z\,f_{0i}(\bs,z)=G_i\,(z\,\xi_i)=G_i\,(T_i^*\,(\xi_i) +\delta_{s_i 0})=
(\bZ_i\,f_{0i})(\bs,z)+\alpha_{i0}(\bs)\,\delta_{s_i 0},
\]
where we have taken into account that
\[
T^n_i(\delta_{s_i 0})=\delta_{s_i-n, 0}=0,\quad \forall n\geq
1,\;\bs\in\Gamma_{II}.
\]
\end{proof}

\subsection{The second system of string equations}

The diagonal solutions
\[
\Phi(z)=\sum_{\alpha} \phi_{\alpha}(z)\,E_{\alpha}
\]
of the condition \eqref{inv} corresponding to the
function $g(z)$ of \eqref{fokg1} are characterized by
\[
\partial_z\,w_i+\phi_0\,w_i-\phi_i\,w_i=0,\quad i=1,\ldots,q.
\]
Hence, setting $\phi_0\equiv 0$ we get
\begin{equation}\label{cov1}
\Phi(z)=\sum_{i} V'(\bc_i,z)\,E_{i}.
\end{equation}
The corresponding covariant derivative is
\begin{equation}\label{cov1}
D_z\,f:=\dfrac{\partial f}{\partial z}-\sum_i
V'(\bc_i,z)\,f\,E_{i},
\end{equation}
so that we may write
\begin{equation}\label{cov2}
D_z(\,E_{0}\,f)=\partial_z f_{00}\,E_{0}+\sum_i (\partial_z
f_{0i}-V'(\bc_i,z)\,f_{0i})\,E_{0i}.
\end{equation}

In order to take advantage of the last identity we observe that \eqref{ste11} can be
generalized to
\begin{equation}\label{pow}
z^n\,f_{0i}(\bs,z)=
(\bZ_i^n\,f_{0i})(\bs,z)-\sum_{r=0}^{n-1}p_{(i,r)}^{(n)}(\bs,z)\,\delta_{s_i+r,
0},\quad \forall \bs\in\Gamma_{II}.
\end{equation}
where the coefficients $p_{(i,r)}^{(n)}(\bs,z)$ are polynomials in
$z$. On the other hand we have that as $z\rightarrow\infty$
\begin{align}\label{mas}
\nonumber &((T_j^*)^n\,f_{0\alpha})(\bs,z)=\begin{cases}
\mathcal{O}\Big(\dfrac{1}{z^n}\Big)\,z^{s_0},\,
\mbox{for $\alpha=0,\,s_j\leq -n$},
\\\\
\mathcal{O}\Big(\dfrac{1}{z}\Big)\,z^{s_i},\,
\mbox{for $\alpha=i\neq j$ and $\,s_j\leq -n$},\\\\
0,\, \mbox{for $s_j>-n$}
\end{cases},\quad  n\geq 1,\\\\
\nonumber &(T_i^n\,f_{0i})(\bs,z)=
\mathcal{O}\Big(\dfrac{1}{z^{n+1}}\Big)\,z^{s_i},\quad n\geq 0.
\end{align}
We are now ready to prove the following result.
\begin{pro}
The function $f$ satisfies the equation
\begin{equation}\label{string2dos}
(D_z+{\cal H})\,(E_{0}\,f)(\bs,z)=\sum_{i=1}^q \Delta_i(\bs,z)\,E_{0i},
\quad \forall \bs\in\Gamma_{II},
\end{equation}
where ${\cal H}$ is the operator
\begin{equation}\label{oper}
\mathcal{H}:=\sum_{j
=1}^q V'(\bc_j,\mathcal{Z}_j)_{(j,+)},
\end{equation}
and  $\Delta_i(\bs,z)$ are functions of the form
\begin{equation}\label{pld}
\Delta_i(\bs,z)=\sum_{n= 1}^{N_i} p_{(i,n)}(\bs,z)\,\delta_{s_i+n,
0}.
\end{equation}
with $p_{(i,n)}(\bs,z)$ being polynomials in $z$ and
$N_i=degree\,V(\bc_i,z)-2$.
\end{pro}
\begin{proof}
From \eqref{cov2} it follows that
\begin{equation}\label{covf22}\everymath{\displaystyle}
\begin{array}{l}
D_z(E_{0}\,f)(\bs,z)+{\cal H}(E_{0}\,f)(\bs,z)\,=
\left[\partial_zf_{00}+\sum_{j=1}^qV'(\bc_j,\mathcal{Z}_j)
_{(j,+)}\,f_{00}\right]\,E_{0}\\  \\
\qquad\qquad +\sum_{i=1}^q\left[\partial_zf_{0i}+ \sum_{j=1}^q V'(\bc_j,\mathcal{Z}_j)_{(j,+)}\,f_{0i}
 -V'(\bc_i,z)\,f_{0i}\right]E_{0i}.
\end{array}\end{equation}
Using  \eqref{mas} we find
\begin{align}\label{s211}
\nonumber &\partial_zf_{00}+\sum_{j=1}^qV'(\bc_j,\mathcal{Z}_j)
_{(j,+)}\,f_{00}=\mathcal{O}\Big(\dfrac{1}{z}\Big)\,z^{s_0},\\
&\partial_zf_{0i}+ \sum_{j\neq i} V'(\bc_j,\mathcal{Z}_j)_{(j,+)}\,f_{0i}=\mathcal{O}\Big(\dfrac{1}{z}\Big)\,z^{s_i},
\\
\nonumber &\Big(V'(\bc_i,\mathcal{Z}_i)_{(i,+)}-V'(\bc_i,z)\Big)\,f_{0i}=
\Big(V'(\bc_i,\mathcal{Z}_i)-V'(\bc_i,z)\Big)\,f_{0i}+\mathcal{O}\Big(\dfrac{1}{z}\Big)\,z^{s_i}.
\end{align}
On the other hand
\eqref{pow} implies
\begin{equation}\label{imp}
\Big(V'(\bc_i,\mathcal{Z}_i)-V'(\bc_i,z)\Big)\,f_{0i}
=\sum_{n\geq 1}n\,c_{in}\,(\mathcal{Z}_i^{n-1}-z^{n-1})\,f_{0i}=\Delta_i(\bs,z)\,f_{0i},
\end{equation}
where
\begin{equation}\label{pld}
 \Delta_i(\bs,z):
=\sum_{n\geq 1}n\,c_{in}\,\sum_{r=0}^{n-2}p_{(i,r)}^{(n-1)}(\bs,z)\,\delta_{s_i+r,
0}.
\end{equation}
Hence Eq.\eqref{covf22} says that
\[
(D_z+{\cal H})\,(E_{0}\,f)(\bs,z)-\sum_{i=1}^q \Delta_i(\bs,z)\,E_{0i}=
{\cal
O}(\frac{1}{z})\,f_0(\bs,z).
\]
The first member
of this equation is a solution of the Riemann-Hilbert problem for all $\bs\in\Gamma_{II}$  so that from Proposition 2
the statement follows.
\end{proof}

As a consequence we deduce the string equations
\begin{teh}
The multiple orthogonal polynomials of type II verify
\begin{equation}\label{string1pol}
\partial_z\,P(\bn,z)+\sum_{j
=1}^q V'(\bc_j,\mathcal{Z}_j)_{(j,+)}\,P(\bn,z)=0.
\end{equation}

\end{teh}

\subsection{Orlov operators}

We define  the Orlov operators $\bM_i$ by
\begin{equation}\label{or2}
\bM_i:=G_i\cdot(s_i-1)\cdot T_i\cdot G_i^{-1}.
\end{equation}
They satisfy $[\bZ_i,\bM_i]=\mathbb{I}$ and can be expanded as
\begin{equation}\label{or2e}
\bM_i=\sum_{n\geq 1}\mu_{in}(\bs)\,T_i^n,
\end{equation}
where
\begin{equation}
\mu_{i1}(\bs)=\dfrac{s_i-1}{v_i(\bs-\be_i)}.
\end{equation}
\begin{pro}
The functions $f_{0i}$
satisfy the equations
\begin{equation}\label{ste2dos}
\partial_z\,f_{0i}(\bs,z)=(\bM_i\,f_{0i})(\bs,z),\quad \forall \bs\in\Gamma_{II}.
\end{equation}
\end{pro}
\begin{proof}
From the definition of $G_i$ we have
\[
\partial_z\,f_{0i}=G_i\,((s_i-1)\,z^{-1}\,\xi_i)=G\cdot(s_i-1)(T_i\,\xi_i) =
G_i\cdot(s_i-1)\cdot T_i\cdot G_i^{-1} f_{0i}=\bM_i\,f_{0i}.
\]
\end{proof}

\section{The large-$\bn$ limit}

The large-$\bn$ limit of multiple orthogonal polynomials is closely connected to the \emph{quasiclassical} limit of the functions $f_{0\alpha}(\bs,z)$. In this section  we will  consider  these functions for large values of the discrete parameters $s_i$
\[
s_i>>1, \forall i\quad (\mbox{Type I case});\quad
s_i<<-1, \forall i\quad (\mbox{Type II case}).
\]
Note that in particular the string equations \eqref{string1} and \eqref{string2dos} simplify since all the $\delta$ terms vanish. As a consequence the resulting  equations  are the same for both types of multiple orthogonal polynomials and can be summarized as follows:

\begin{equation}\label{type}
\begin{cases}
z\,f_{0\alpha}=\Big(T_{i}+u_{i}(\bs)+\sum_{j}\,v_j(\bs)\, T_j^*\Big)\,f_{0\alpha},\quad \forall \alpha,i;
\\\\
\partial_z\,f_{00}=-{\cal H}\,f_{00},\quad
\partial_z\,f_{0i}=\Big(-{\cal H}+V'(\bc_i,\bZ_i)\Big)\,f_{0i}.
\end{cases}
\end{equation}

 In order to define the large-$\bn$ limit we introduce an small parameter $\epsilon$, define \emph{slow variables}
\begin{equation}\label{slow}
t_{i}:=\epsilon\,s_{i},;\quad t_0:=-\sum_{i=1}^q t_i,\quad \bt:=(t_1,\ldots,t_q),
\end{equation}
and rescale the exponents of the weight functions \eqref{gtypeI} and \eqref{fokg1} as
\[
w_i(\epsilon,z)=\exp\Big(\mp\,\dfrac{V(\bc_i,z)}{\epsilon}\Big),
\]
where the exponent sign is negative (positive) for polynomials of type I (type II).
Moreover, we perform  a continuum limit in which as $\epsilon\rightarrow 0$, the discrete parameters $s_i$ tend to $+\infty$ ($-\infty$) for the type I case (type II case) and $t_{\alpha}$ become continuous variables.

\vspace{0.3cm}

  The problem now is to determine solutions $f_{0\alpha}(\epsilon,\bt,z)$ of \eqref{type} defined for $\bt$ on some domain $\Omega$ of $\R^q$, that have the \emph{quasiclassical} form \cite{tak2}
\begin{equation}\label{quass}
f_{0\alpha}(\epsilon,\bt,z)=z^{\delta_{0\alpha}-1}\,\exp{\Big(\dfrac{1}{\epsilon}\,\mathbb{S}_{\alpha}\Big)},\quad
\mathbb{S}_{\alpha}
=t_{\alpha}\,\log{z}+\sum_{n\geq 0}\dfrac{1}{z^n}\mathbb{S}_{\alpha n},
\end{equation}
where
\[
\mathbb{S}_{\alpha n}=\sum_{k\geq 0}\epsilon^k\,\mathbb{S}_{\alpha n}^{(k)}(\bt),\quad n\geq 0;\quad
\mathbb{S}_{0 0}\equiv 0.
\]
Note the leading behaviour
\begin{equation}\label{class}
f_{0\alpha}(\epsilon,\bt,z)=z^{\delta_{0\alpha}-1}\,
\exp{\Big(\dfrac{1}{\epsilon}\,S_{\alpha}+\mathcal{O}(1)\Big)},\quad
\mbox{as $\epsilon \rightarrow 0$},
\end{equation}
where
\begin{equation}\label{act}
S_{\alpha}(\bt,z):=t_{\alpha}\,\log{z}+\sum_{n\geq 0}\dfrac{1}{z^n}\,S_{\alpha n}(\bt),\quad
S_{\alpha n}=\mathbb{S}_{\alpha n}^{(0)},\quad S_{00}\equiv 0,
\end{equation}
are the \emph{classical action} functions.

\vspace{0.3cm}

In terms of  slow variables the operators
 $T_i$ and $T_i^*$ become translation operators  \begin{equation}\label{tes}
T_i=\exp(-\epsilon\,\partial_{i}),\quad
T_i^*=T_i^{-1}=\exp(\epsilon\,\partial_{i}),\quad \partial_{i}:=\dfrac{\partial\,}{\partial\,t_i}.
\end{equation}
Hence, we have the following useful relations
\begin{equation}\label{cuas}
T_i^{\pm 1}\,f_{0\alpha}=\exp\Big(\mp \,\partial_{i}\,S_{\alpha}+
\mathcal{O}(\epsilon)\Big)\,f_{0\alpha}.
\end{equation}
It is now a simple matter to deal with the corresponding  dressing and Lax-Orlov operators. Indeed, expressing the functions  \eqref{quass} in the form
\[
f_{0\mu}=\Big(\delta_{0\mu}+\sum_{n\geq 1}\dfrac{\alpha_{\mu n}(\epsilon,\bt)}{z^n}\Big)\,\exp\Big(\dfrac{t_{\mu}}{\epsilon}\,\log z\Big),
\]
we have
\begin{align*}
f_{0i}&=G_{i}\,\exp\Big(\dfrac{t_i}{\epsilon}\,\log z\Big),\quad
G_{i}:=\sum_{n\geq 1}\alpha_{in}(\epsilon,\bt)\,T_i^n\\\\
\bZ_i:&=G_{i}\,T_i^{-1}\,G_{i}^{-1},\quad
\bM_i:=G_{i}\cdot t_i\cdot T_i\cdot\,G_{i}^{-1}
\end{align*}
We can also introduce Lax-Orlov associated with $f_{00}$. In fact we may do it in $q$ different ways
\begin{align*}
f_{00}&=G_0^{(i)}\,\exp\Big(\dfrac{t_0}{\epsilon}\,\log z\Big),\quad G_0^{(i)}=1+\sum_{n\geq 1}\alpha_{0n}(\epsilon,\bt)\,T_i^{-n},\\ \\
\bZ_0^{(i)}:&=G_0^{(i)}\,T_i\,(G_0^{(i)})^{-1},\quad
\bM_0^{(i)}:=G_0^{(i)}\cdot t_0\cdot T_i^{-1}\cdot\,(G_0^{(i)})^{-1}.
\end{align*}

In terms of Lax-Orlov operators and taking into account the assumption \eqref{quass} the  system of string equations \eqref{type} becomes
\begin{equation}\everymath{\displaystyle}\label{typeb}
\begin{cases}
z\,f_{0\alpha}=\bZ_{\alpha}\,f_{0\alpha}=\Big(T_{j}+\mathrm{u}_j(\epsilon,\bt)+\sum_{k}\,\mathrm{v}_k(\epsilon,\bt)\, T_k^*\Big)\,f_{0\alpha},\quad \forall \alpha,j;
\\\\
\epsilon\,\partial_z\,f_{00}=\bM_0\,f_{00}=-{\cal H}\,f_{00},\quad
\epsilon\,\partial_z\,f_{0j}=\bM_j\,f_{0j}=\Big(-{\cal H}+V'(\bc_j,\bZ_j)\Big)\,f_{0i},
\end{cases}
\end{equation}
for all choices  $\bZ_0=\bZ_0^{(i)},\bM_0= \bM_0^{(i)}$. It follows from \eqref{quass}
that the recurrence coefficients $\mathrm{u}_j$ and  $\mathrm{v}_j$ can be written as quasiclassical expansions of the form
\begin{equation}\label{uv}
\mathrm{u}_i=u_i(\bt)+\sum_{n=1}^{\infty}\epsilon^n\,u_{i,n}(\bt)
,\quad
\mathrm{v}_i=v_i(\bt)+\sum_{n=1}^{\infty}\epsilon^n\,v_{i,n}(\bt),
\end{equation}

\subsection{Leading behaviour and hodograph equations}

Our next aim is to characterize the leading behaviour of the solutions $f_{0\alpha}$ of \eqref{typeb}. More concretely we are going to see how the leading terms
$$
\bu:=(u_1(\bt),\ldots,u_q(\bt)),\quad \bv:=(v_1(\bt),\ldots,v_q(\bt)),
$$
of the recurrence coefficients \eqref{uv} are determined by  a system of hodograph type equations.

In order to formulate the classical limits $(z_{\alpha},m_{\alpha})$ of the Lax-Orlov operators $(\bZ_{\alpha},\bM_{\alpha})$ we observe that as a consequence of the first group of string equations in \eqref{typeb} we have that
\begin{equation}\label{con}
\Big(T_{i}+\mathrm{u}_i(\epsilon,\bt)\Big)\,f_{0\alpha}=\Big(T_{j}+
\mathrm{u}_j(\epsilon,\bt)\Big)\,f_{0\alpha},\quad \forall i,j,\alpha.
\end{equation}
Then, using \eqref{cuas} we get
\begin{equation}\label{ind}
\exp\Big(-\partial_i\,S_{\alpha}(\bt,z)\Big)+u_i(\bt)=
\exp\Big(-\partial_j\,S_{\alpha}(\bt,z)\Big)+u_j(\bt),\quad \forall i,j.
\end{equation}
In view of these identities we define $z_{\alpha}(\bt,p)$ by the implicit equations
\begin{equation}\label{zet}
p=\exp\Big(-\partial_{i}\,S_{\alpha}(\bt,z_{\alpha}(\bt,p))\Big)+u_i(\bt).
\end{equation}
Notice that according to \eqref{ind} these definitions are independent of the value of the index $i$ used in \eqref{zet}. Moreover, \eqref{zet} imply
\begin{equation}\label{di}
\partial_{i}\,S_{\alpha}(\bt,z_{\alpha})=-\log(p-u_i(\bt)).
\end{equation}
From the asymptotic expansion \eqref{act} of the action functions $S_{\alpha}$ and the defining equations \eqref{zet}  it is straightforward to prove that the Lax functions can be expanded as
\begin{equation}\everymath{\displaystyle}\label{cetas}
\begin{cases}
z_0=p+\sum_{n=1}^{\infty}
\dfrac{v_{0n}(\bt)}{p^n},\quad p\rightarrow \infty,\\\\
z_i=\dfrac{v_{i}(\bt)}{p-u_i(\bt)}+
\sum_{n=0}^{\infty}v_{in}(\bt)\,(p-u_i(\bt))^n,\quad  p\rightarrow u_i(\bt)
.
\end{cases}
\end{equation}
On the other hand, we define the corresponding  Orlov functions $m_{\alpha}(\bt,z_{\alpha})$ by
\begin{equation}\label{orll}
m_{\alpha}(\bt,z_{\alpha}):=\partial_z\, S_{\alpha}(\bt,z_{\alpha}).
\end{equation}
The definitions  \eqref{zet} and \eqref{orll} provide the classical limits of the Lax-Orlov operators. Indeed, from \eqref{typeb} it follows at once that
\begin{align}\label{quas}
\nonumber (\bZ_{\alpha}\,f_{0\alpha})(\bt,z_{\alpha}(\bt,p))&=
z_{\alpha}(\bt,p)\,f_{0\alpha}(\bt,z_{\alpha}(\bt,p)),\\\\
\nonumber(\bM_{\alpha}\,f_{0\alpha})(\bt,z_{\alpha}(\bt,p))&=
\Big(m_{\alpha}+\mathcal{O}(\epsilon)\Big)\,f_{0\alpha}(\bt,z_{\alpha}(\bt,p)),
\end{align}
for all choices of $\bZ_0=\bZ_0^{(i)}, \bM_0=\bM_0^{(i)}$. In particular this means that all the pairs of Lax-Orlov operators $(\bZ_0^{(i)}, \bM_0^{(i)})$ have the same classical limit given by $(z_0(\bt,p), m_0(\bt,p))$.
\begin{teh}
The Lax-Orlov functions satisfy the classical string equations
\begin{equation}\label{cstt}
\begin{cases}
z_0=z_1=\cdots=z_q=E(\bu,\bv,p),
\\\\
 m_0=m_1-
V'(\bc_1,z_1)=\cdots=m_q-\,V'(\bc_q,z_q)= -H(\bu,\bv,p),
\end{cases}
\end{equation}
where
\begin{equation}\label{eh}
E:=p+\sum_{k=1}^q
\dfrac{v_k(\bt)}{p-u_k(\bt)},\quad  H:=\sum_{k=1}^q V'(\bc_k,z_k)_{(k,+)},
\end{equation}
and $(\quad)_{(k,+)}$  stand for the
projections of power series in $(p-u_k)^n,\,(n\in\Z)$ on the subspaces generated by
$(p-u_k)^{-n}\, (n\geq 1)$.

\end{teh}
\begin{proof}
Taking into account that
\[
T_j^n\,f_{0\alpha}(\bt,z_{\alpha}(p,\bt))=\Big((p-u_j)^n+\mathcal{O}(\epsilon)\Big)\,
f_{0\alpha}(\bt,z_{\alpha}(\bt,p)),\quad n=\pm 1,\pm 2,\ldots,
\]
it is easy to see that the equations \eqref{cstt} are the classical limit ($\epsilon\rightarrow 0$) of the system \eqref{typeb}.
\end{proof}

 In view of the first group of equations in \eqref{cstt}, it is clear that the functions $\bu$ and $\bv$ are the only unknowns  for determining the Lax-Orlov functions. However, the Lax-Orlov functions must verify the correct asymptotic expansions. Obviously, the functions $z_{\alpha}=E$ satisfy \eqref{cetas}. Nevertheless, Eq.\eqref{act} requires  the Orlov functions to satisfy
\begin{equation}\label{asm}
m_{\alpha}=\dfrac{t_{\alpha}}{z_{\alpha}}-\sum_{n\geq 1}
\dfrac{n\,S_{\alpha n}(\bt)}{z_{\alpha}^{n+1}},\quad \mbox{as
$z_{\alpha}\rightarrow\infty$},
\end{equation}
and this behaviour must be compatible with the second group of  equations in \eqref{cstt}
\begin{equation}\label{asm1}
m_0=-H(\bu,\bv,p),\quad m_i
=V'(\bc_i,E)-H(\bu,\bv,p),
\end{equation}
where we have already inserted the substitutions $z_i=E$.
Let us analyze the equations \eqref{asm1} in terms of series expansions as $p \rightarrow\infty$ for $m_0$, and as
$p\rightarrow u_i(\bt)$ for $m_i$. If we take into account that  \[
\dfrac{1}{z_0}=\dfrac{1}{p}+\mathcal{O}\Big(\dfrac{1}{p^2}\Big),\quad
H=\mathcal{O}\Big(\dfrac{1}{p}\Big),\quad 
p \rightarrow\infty,
\]
\[
\dfrac{1}{z_i}=\mathcal{O}((p-u_i)),
\quad
\dfrac{1}{p-u_j}=\mathcal{O}(1);\quad
V'(\bc_i,E)-H=\mathcal{O}(1)
,\quad j\neq i,\quad p \rightarrow u_i(\bt),
\]
then the consistency between  \eqref{asm1} and \eqref{asm} requires 
\begin{equation}\everymath{\displaystyle}\label{hodo1}
\oint_{\gamma_0}\dfrac{\d p}{2i\pi}\, H(\bu,\bv,p)=-t_0,
\end{equation}
and
\begin{equation}\everymath{\displaystyle}\label{hodo2}
\begin{cases}
\oint_{\gamma_i}\dfrac{\d p}{2i\pi}\,\dfrac{V'(\bc_i,E(\bu,\bv,p))-H(\bu,\bv,p)}{p-u_i}=0
 \\\\
 \oint_{\gamma_i}\dfrac{\d p}{2i\pi}\,\dfrac{V'(\bc_i,E(\bu,\bv,p))-H(\bu,\bv,p)}{(p-u_i)^2}=\dfrac{t_i}{v_i}.
\end{cases}
\end{equation}
These conditions are obtained
by comparing coefficients of $p^{-1}$ and $(p-u_i(\bt))^n$ with $(n=0,1)$ in the equations \eqref{asm1} for $m_0$ and $m_i$, respectively.
Here $\gamma_i$ are positively oriented small circles around $p=u_i$ such that  $p=u_j$ is outside $\gamma_i$ for
all $j\neq i$,  and $\gamma_0$ is a large positively oriented circles which
encircles all the $\gamma_i $ (see fig.1).

 Identifying the coefficients of the remaining powers $p^{-n}$ and $(p-u_i(\bt))^n$ in \eqref{asm1} determines the Orlov functions in terms of $(\bu,\bv)$.

\begin{figure}\label{fig 1}
\begin{center}
\psfrag{p}{$p$-plane}\psfrag{z}{$z_\mu$-plane}
\psfrag{a}{$\gamma_0$}\psfrag{b}{$\gamma_1$}\psfrag{c}{$\gamma_2$}
\psfrag{d}{$\gamma_3$}
\includegraphics[width=6cm]{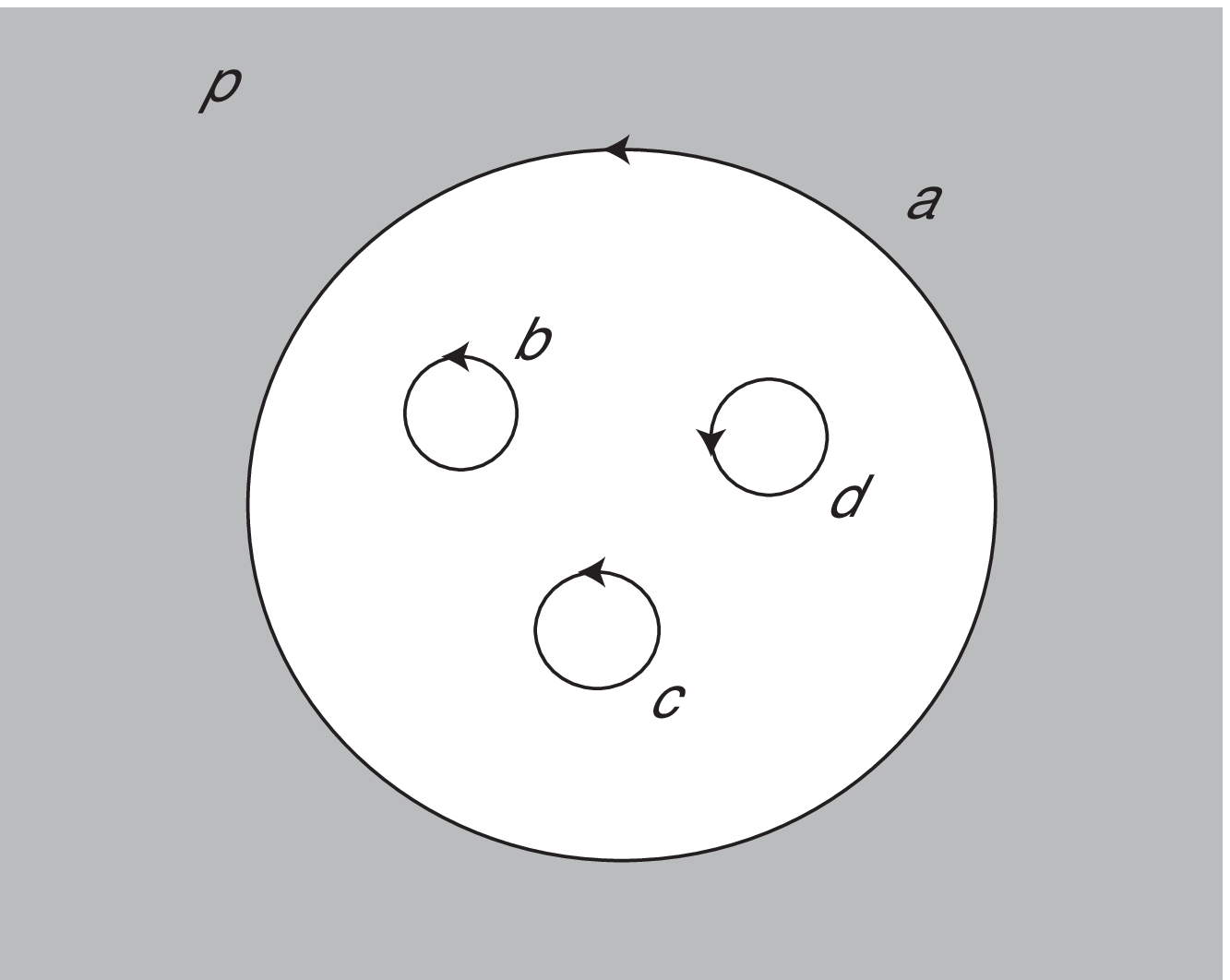}
\end{center}\caption{}
\end{figure}

 The equation \eqref{hodo1}  is a consequence of the second group of equations in \eqref{hodo2} and the fact that $t_0:=-\sum_i t_i$. To see this, notice that
\begin{align*}\everymath{\displaystyle}
\oint_{\gamma_0}\, H(p)\,\d p&=\oint_{\gamma_0}\, H(p)\,\partial_p\,E(p)\,\d p,\\\\
\oint_{\gamma_i}\,(V'(\bc_i,E(p))-H(p))\,\dfrac{v_i}{(p-u_i)^2}\,\d p&=-
\oint_{\gamma_i}\,(V'(\bc_i,E(p))-H(p))\,\partial_p\,E(p)\,\d p.
\end{align*}
Hence

\begin{align*}
&-\oint_{\gamma_0}\, H(p)\,\d p+\sum_i
\oint_{\gamma_i}\,(V'(\bc_i,E(p))-H(p))\,\dfrac{v_i}{(p-u_i)^2}\,\d p
\\\\
&=-\oint_{\gamma_0}H(p)\,\partial_p\,E(p)\,
\d p-
\sum_{i}\oint_{\gamma_i}\Big(V'(\bc_i,E(p))-H(p)\Big)\,\,\partial_p\,E(p)
\,\d p
\\\\
&=-\oint_{\gamma_0-\sum_{i}\gamma_{i}}\,H(p)\,\partial_p\,E(p)\,\d p-
\sum_{i}\oint_{\gamma_i}V'(\bc_i,E(p))\,\partial_p\,E(p)\,\d p=0,
\end{align*}
where we have taken into account that $V'(\bc_i,E(p))\,\partial_p\,E(p)=
\partial_p\,\Big(V(\bc_i,E(p))\Big)$. Moreover,
$H(p)\,\partial_p\,E(p)$ is a rational function of $p$ with poles at the points $p_i=u_i$ only and
\[
\gamma_0-\sum_{i}\gamma_{i}\sim 0 \quad \mbox{in}\quad
\C\setminus\{p_1,\dots,p_q\}.
\]
Therefore we are finally lead to the system \eqref{hodo2}  of $2q$
equations for determining the $2q$ functions $u_i,v_i$. These equations are of hodograph type as they depend linearly on the
parameters $\bt$ and $\bc_i$. For example the first few terms are
\begin{equation}\everymath{\displaystyle}\label{hodo2a}
\begin{cases}
c_{i1}+2\,c_{i2}\,u_i+\sum_{j\neq i}\dfrac{(c_{i2}-c_{j2})\,v_j}{u_i-u_j}+\cdots=0
 \\\\
2\,c_{i2}-\sum_{j\neq i}\dfrac{(c_{i2}-c_{j2})\,v_j}{(u_i-u_j)^2}+\cdots =\dfrac{t_i}{v_i}.
\end{cases}
\end{equation}

\subsection{Connection with the Whitham hierarchy}

If we assume that the  coefficients $\bc_i$ of exponents of the weight functions \eqref{gtypeI} and \eqref{fokg1} are free parameters and write them in the form
\begin{equation}\label{kpp}
\bc_i=\bt_0-\bt_i,\quad \bt_{\alpha}=(t_{\alpha 1},\ldots,t_{\alpha n},\ldots)\in \C^{\infty},
\end{equation}
then, as we are going to see,  the solution of \eqref{cstt} turns out to determine a solution of the Whitham hierarchy of dispersionless integrable system
\cite{krich}.

Let us introduce the modified  Orlov functions
\begin{equation}\label{asmn}
\widetilde{m}_{\alpha}=V'(\bt_{\alpha},z_{\alpha})+m_{\alpha}.
\end{equation}
It is clear that $(z_{\alpha},\widetilde{m}_{\alpha})$ solve
the system
\begin{equation}\label{csttn}
\begin{cases}
z_0=z_1=\cdots=z_q,
\\\\
 \widetilde{m}_0=\widetilde{m}_1=\cdots=\widetilde{m}_q.
\end{cases}
\end{equation}
Moreover, they are rational functions of $p$ with poles  at the points $p_i=u_i$ only. Furthermore, they satisfy the asymptotic
properties \eqref{cetas} and
\[
\widetilde{m}_{\alpha}
=\sum_{n\geq 1}\,n\,t_{\alpha n}\,z_{\alpha}^{n-1}+
\dfrac{t_{\alpha}}{z_{\alpha}}-\sum_{n\geq 1}
\dfrac{n\,S_{\alpha n}(\bt)}{z_{\alpha}^{n+1}},\quad \mbox{as
$z_{\alpha}\rightarrow\infty$}.
\]
Thus, the functions $(z_{\alpha},\widetilde{m}_{\alpha})$ satisfy all the conditions of Theorem 1 of \cite{mio00} and, as a consequence, they verify the equations of the Whitham hierarchy
\begin{equation}\label{wh} \frac{\partial z_{\alpha}}{\partial
t_{\mu n}}=\{\Omega_{\mu n}, z_{\alpha}\},\quad
\frac{\partial \widetilde{m}_{\alpha}}{\partial
t_{\mu n}}=\{\Omega_{\mu n}, \widetilde{m}_{\alpha}\}
\end{equation}
where the Poisson bracket is given by
\[
\{F,G\}:=\frac{\partial F}{\partial p} \frac{\partial G}{\partial
x} -\frac{\partial F}{\partial x}\frac{\partial G}{\partial p}, \quad x:=t_{01}.
\]
and the Hamiltonian functions are
\begin{gather}\label{1.3}\everymath{\displaystyle}
\Omega_{\mu n}:=\begin{cases}   (z_\mu^n)_{(\mu,+)} ,& n\geq 1
,\\\\
-\log(p-u_i), &n=0,\quad \mu=i=1,\dots,q.
\end{cases}
\end{gather}
Here $(\cdot)_{(0,+)}$ stands for the
projector on  $\{p^n\}_{n=0}^\infty$ .

\vspace{0.3cm}

In this way we conclude that $(z_{\alpha},\widetilde{m}_{\alpha})$, as functions of the coupling constants $\bc_i=\bt_0-\bt_i$, determine a reduced solution of the Whitham hierarchy. This property is in complete agreement with the results of recents works \cite{tak2} which  prove that the universal Whitham hierarchy can be obtained as a particular dispersionless limit of the multi-component KP hierarchy. Moreover, as it as been observed in \cite{ad2}, appropriate deformations of the Riemann-Hilbert problems for multiple orthogonal polynomials determine solutions of the multi-component KP hierarchy. In fact these deformations correspond to   the flows induced by changes in the parameters $\bt_{\alpha}$. Indeed, for both types of multiple orthogonal polynomials, \eqref{kpp} implies
\[
\partial_{t_{\alpha n}}\,g=[z^n\,E_{\alpha},g].
\]
Therefore the covariant derivatives
\[
D_{\alpha n}\,f:=\partial_{\alpha n}\,f+z^n\,f\,E_{\alpha},
\]
are symmetries of the corresponding Riemman-Hilbert problems. Hence,
using Proposition 2 one concludes that
\begin{equation}\label{kpls}
\partial_{\alpha n}\,f+(z^n\,f\,E_{\alpha}\,f^{-1})_-\,f=0,
\end{equation}
where $(\quad)_-$  stands for the
projections of power series in $z^k,\,(k\in\Z)$ on the subspaces generated by
$z^{-k}\, (k\geq 1)$. The equations \eqref{kpls} constitute the linear system of the multi-component KP hierarchy.

\section{Applications: random matrix models and \\ non-intersecting Brownian
motions}

As we have seen the   multiple orthogonal polynomials of type I are the elements $f_{0i}$ of the fundamental solution of their associated RH problem. Thus, in the quasiclassical limit we have
\[
A_i(\bn,z)\sim \dfrac{1}{z}
\exp{\Big(\dfrac{1}{\epsilon}\,S_{i}(\bt,z))}\Big),\quad
\mbox{as $\epsilon \rightarrow 0$},
\]
where $S_{i}=S_{i}(\bt,z)$  are the classical action functions
defined in \eqref{act}. Hence
\begin{equation}\label{n1}
\epsilon\,\partial_z\,\log A_i(\bn,z)\sim \partial_z\,S_i(\bt,z)-\dfrac{\epsilon}{z}=m_i(\bt,z)-\dfrac{\epsilon}{z}.
\end{equation}
On the other hand, if we denote by $x_i$ the roots of $A_i(\bn,z)$ we have
\[
\partial_z\,\log A_i(\bn,z)=\,\sum_{i=1}^{n_i-1}
\dfrac{1}{z-x_i}.
\]
Thus if we assume that in the large-$\bn$ limit the roots of $A_i$ are distributed with a continuous density $\rho_i=\rho_i(x)$ on some compact (possibly disconnected) support $I_i\subset \R$
\begin{equation}\label{n2}
\epsilon\,\sum_{i=1}^{n_i-1}
\dfrac{1}{z-x_i}\sim \int_{I_i}\dfrac{\rho_i(x)}{z-x}\,\d x ,\quad
\mbox{as $\epsilon \rightarrow 0$}.
\end{equation}
from \eqref{n1} and \eqref{n2} we deduce the important relation
\begin{equation}\label{n3}
m_i(z)= \int_{I_i}\dfrac{\rho_i(x)}{z-x}\,\d x,
\end{equation}
where $m_i$, $I_i$ and $\rho_i$ depend on the slow variables $\bt$. This means that  the Orlov functions $m_i$ are the Cauchy transforms of the root densities $\rho_i$. Hence, they determines the distribution of roots in the large-$\bn$ limit according to
\begin{equation}\label{n4}
m_{i+}(x)-m_{i-}(x)=-2\,i\pi\,\rho_i(x),\quad x\in I_i.
\end{equation}
Moreover, from \eqref{asm} we see that
\begin{equation}\label{n5}
\int_{I}\rho_i(x)\,\d x=t_i.
\end{equation}

\vspace{0.3cm}

 On the other hand, the multiple orthogonal polynomials of type II represent the element  $f_{00}$ of their associated RH problem. Therefore, in the quasiclassical limit we have
\[
P(\bn,z)\sim
\exp{\Big(\dfrac{1}{\epsilon}\,S_{0}(\bt,z)}\Big),\quad
\mbox{as $\epsilon \rightarrow 0$},
\]
Thus if we assume that in the large-$\bn$ limit the roots of $P(\bn,z)$ tend to be distributed with a continuous density $\rho_0=\rho_0(x)$ on some compact support $I_0\subset \R$,
 we deduce
\begin{equation}\label{n30}
m_0(z)= \int_{I_0}\dfrac{\rho_0(x)}{z-x}\,\d x,
\end{equation}
where $m_0$, $I_0$ and $\rho_0$ depend on the slow variables $\bt$. Thus the Orlov function $m_0$ is the Cauchy transform of the density $\rho_0$ and therefore
\begin{equation}\label{n40}
m_{0+}(x)-m_{0-}(x)=-2\,i\pi\,\rho_0(x),\quad x\in I_0.
\end{equation}
Note also that
\begin{equation}\label{n50}
\int_{I_0}\rho_0(x)\,\d x=t_0.
\end{equation}

\vspace{0.3cm}

The string equations  \eqref{cstt} provide also useful information to
determine the limiting supports and the root densities.
They imply
\[
m_0(z)= -H(p_0(z)),\quad m_i(z)=V'(\bc_i,z) -H(p_i(z)),
\]
where $p_{\alpha}(z)$  denote the $q+1$ inverses of the map
\[
 z(p):=E(p)
=p+\sum_{k=1}^q
\dfrac{v_k(\bt)}{p-u_k(\bt)},
\]
verifying
\[
p_0(z)=z+\mathcal{O}\Big(\dfrac{1}{z}\Big),\quad
p_i(z)=u_i+\mathcal{O}\Big(\dfrac{1}{z}\Big);\quad \mbox{as $z\rightarrow\infty$}.
\]
Therefore \eqref{n4} and \eqref{n40} reduce to
\begin{equation}\label{n4hg}
H(p_{\alpha+}(x))-H(p_{\alpha-}(x))=2\,i\pi\,\rho_{\alpha}(x),\quad x\in I_{\alpha}.
\end{equation}
In general the limiting supports $I_{\alpha}$ may consist of several disconnected segments
 \[
\quad I_{\alpha} = \bigcup_{k=1}^{d_{\alpha}}
I_{\alpha k}
\]
which, due to \eqref{n4hg},  constitute the branch cuts of the  functions $H(p_{\alpha}(z))$ . As a consequence the end-points of the segments $I_{\alpha k}$  are the branch points of these functions, which are in turn given by the critical points $x_i$ of the function $z(p)=E(p)$
\begin{equation}\label{n6}
x_i=E(q_i)\in\R,\quad \partial_p E(q_i)=0.
\end{equation}

\subsection{The Hermitian matrix model}

For $q=1$ the multiple orthogonal polynomials of type II reduce to the orthogonal polynomials on the real line associated to the weight function $w=\exp V(\bc,z)$. These polynomials are connected to the  random matrix model of $n {\times} n$ Hermitian matrices \cite{gin}-\cite{dei}
\begin{equation}\label{h10}
Z_n=\int \d M \exp\Big(\mbox{Tr}\,V(\bc,M)\Big),
\end{equation}
through the crucial relation
\begin{equation}\label{h2}
P_n(z)=\mathbb{E}\,[\mbox{det}(z-M)],
\end{equation}
where $\mathbb{E}$ denotes the expectation value with respect to the probability measure determined by \eqref{h10}. This means that in the large-$n$ limit the eigenvales of $M$ converge with unit probability to the roots of $P_n$. As a consequence the root density $\rho_0$ of the family of polynomials represents the eigenvalue density of the matrix model.

 The Hermitian matrix model provides an appropriate  example to illustrate all the aspects of our method for characterizing the quasiclassical limit. In this case we set $\epsilon:=1/n,\,t_0=1$ and we have
\[
z(p)=E(u,v,p)=p+\dfrac{v}{p-u}.
\]
Here $u$ and $v$ depend on the coupling constants
$\bc=(c_1,c_2,\ldots)$ and can be determined by means of the hodograph equations \eqref{hodo1}-\eqref{hodo2}
\[
\everymath{\displaystyle}
\oint_{\gamma_0}\dfrac{\d p}{2i\pi}\, H(p)=-1,
\quad
\oint_{\gamma_1}\dfrac{\d p}{2i\pi}\,\dfrac{V'(\bc,E(p))-H(p)}{p-u}=0.
\]
By introducing the change of variable $p-u\rightarrow p$ these equations are equivalent to the well-known system \cite{gin} \begin{equation}\label{gin}
\oint_{\gamma}\dfrac{\d p}{2i\pi}\,V'(\bc,\,p+u+\dfrac{v}{p})=-1,\quad
\oint_{\gamma}\dfrac{\d p}{2i\pi\,p}\,V'(\bc,\,p+u+\dfrac{v}{p})=0,
\end{equation}
which characterizes the \emph{spherical limit} in the Hermitian matrix model of 2D gravity.
Here $\gamma$ is a large  positively oriented circle around the origin.

The critical points of $E$ are
$q_{\pm}=u\pm\sqrt{v}$, so that the support of eigenvalues is
\begin{equation}\label{sup}
I=[x_-,x_+],\quad x_{\pm}:=u\pm 2\,\sqrt{v}.
\end{equation}
We use \eqref{n4hg} to determine the density of eigenvalues   according to
\begin{equation}\label{n4h}
H(p_{0+}(x))-H(p_{0-}(x))=2\,i\pi\,\rho_{0}(x),\quad x\in [x_-,x_+].
\end{equation}
Furthermore, the two  inverses
of $z(p)$ are
\begin{equation}\label{bra}
p_{0}(z):=\dfrac{1}{2}(z+u+\sqrt{(z-x_-)(z-x_+)}\,),\quad
p_{1}(z):=\dfrac{1}{2}(z+u-\sqrt{(z-x_-)(z-x_+)}\,)
\end{equation}
and we have
\[
H(p_{0}(z))=V'(\bc,z(p))_{(1,+)}\Big|_{p=p_0(z)}.
\]
Now, using the identities
\[
\dfrac{v}{p-u}=z-p,\quad p^2=(u+z)\,p-z\,u-v,
\]
it is clear that there exist polynomials $\alpha_k(z)$ and
$\beta_k(z)$ satisfying
\begin{equation}\label{idd}
\Big(z(p)^k\Big)_{(1,+)}\Big|_{p=p_0(z)}=\alpha_k(z)+
\beta_k(z)\,p_{0}(z),\quad \Big(z(p)^k\Big)_{(1,+)}\Big|_{p=p_1(z)}=\alpha_k(z)+
\beta_k(z)\,p_{1}(z).
\end{equation}
In particular, taking into account that
\[
p_0(z)=z+\mathcal{O}\Big(\dfrac{1}{z}\Big),\quad
p_1(z)=u+\mathcal{O}\Big(\dfrac{1}{z}\Big);\quad \mbox{as $z\rightarrow\infty$},
\]
from \eqref{idd} we deduce
\begin{equation}\label{bet}
\beta_k(z)=-\Big(\dfrac{z^k}{p_0-p_1}\Big)_{\oplus}=-\Big(\dfrac{z^k}{\sqrt{(z-x_-)(z-x_+)}}\Big)_{\oplus},
\end{equation}
where $(\quad)_{\oplus}$  means the projection of power series in
$z^n,\,(n\in\Z)$ on the subspace generated by $z^{n},\, (n\geq 0)$.
Hence it follows that
\[
H(p_0(z))=\sum_{k\geq
1}k\,c_k\,\Big(\alpha_{k-1}(z)+\beta_{k-1}(z)\,p_0(z)\Big),
\]
and therefore we get
\begin{equation}\label{den}
\nonumber\rho(x)=\dfrac{1}{2\,i\pi}\sum_{k\geq
1}k c_k\beta_{k-1}(x)(p_{0+}(x)-p_{0-}(x))
=-\dfrac{1}{2\,\pi}\Big(\dfrac{V'(\bc,x)}{\sqrt{(x-x_-)(x-x_+)}}\Big)_{\oplus}
\,\sqrt{(x-x_-)(x_+-x)},
\end{equation}
which represents the well-known eigenvalue density for the Hermitian model in the one-cut case.
\subsection{Gaussian models with an external source and non-intersecting Brownian motions}

For $q>1$ the multiple orthogonal polynomials of type II are
connected to the Gaussian Hermitian matrix model with an external
source term $A\,M$ \cite{ortb1}-\cite{ortb4}, where $A$ is a fixed diagonal
$n {\times} n$ real matrix. The partition function of this model is given
by
\begin{equation}\label{h1}
Z_n=\int \d M
\exp\Big(-\mbox{Tr}\,\Big(\dfrac{1}{2}M^2-A\,M\Big)\Big).
\end{equation}
It turns out that if  the eigenvalues of $A$ are given by $a_j
,\,( j = 1,\ldots, q)$ with  multiplicities $n_j$, then the
expectation values
\begin{equation}\label{hb2}
P(\bn,z)=\mathbb{E}\,[\mbox{det}(z-M)],\quad \bn:=(n_1,\ldots,n_q),
\end{equation}
are  multiple orthogonal polynomials with respect to the Gaussian weights
\[
w_j(x) = \exp(a_j\,x-\dfrac{1}{2}\,x^2).
\]
These matrix models are deeply connected to one-dimensional non-intersecting Brownian motion \cite{bmo1}-\cite{bmo3}. More concretely,  the  joint probability density for
the eigenvalues $(\lambda_1,\ldots,\lambda_n)$ of M is the same as
the probability density at time $t\in(0,1)$ for the positions
$(x_1,\ldots,x_n)$ of $n$ non-intersecting Brownian motions starting
at the origin at $t=0$ and forming $q$ groups  ending at $q$ fixed points $b_i,\,(i=1,\ldots,q)$ at $t=1$.
The corresponding dictionary for this duality is
\[
\lambda_j=\dfrac{x_j}{\sqrt{t(1-t)}},\quad
a_k=b_k\,\sqrt{\dfrac{t}{1-t}}.
\]

We discuss next an  example of application to the large-$\bn$ limit of non-intersecting Brownian motions. Let us consider an even number $n$ non-intersecting Brownian motions  ending  at two points $\pm b$
with $n_1=n_2=n/2$ \cite{ortb4}. In this case the slow variables take the values $t_1=t_2=-1/2$. Moreover, we have
$$V(\bc_1,z)=a\,z-\frac{z^2}{2},\quad V(\bc_2,z)=-a\,z-\frac{z^2}{2},\quad a:=b\,\sqrt{\frac{t}{1-t}} ,$$
and
\begin{equation}\label{impo}
z(p)=E(p)=p+\frac{v_1}{p-u_1}+\frac{v_2}{p-u_2},\quad H(p)=-\frac{v_1}{p-u_1}-\frac{v_2}{p-u_2}=p-z(p).
\end{equation}
Using the hodograph equations \eqref{hodo2a} one finds
$$
u_1=a,\quad u_2=-a,\quad v_1=v_2=\frac{1}{2},
$$
so that
$$
z(p)=\frac{p^3+(1-a^2)p}{p^2-a^2}.
$$
The corresponding algebraic function $p=p(z)$ satisfies the Pastur equation \cite{pas}
$$p^3-z\,p^2+(1-a^2)\,p+a^2\,z=0,$$
which defines a three-sheeted Riemann surface.
The restrictions of $p(z)$ to the three sheets are the functions $p_{\alpha}(z)$ characterized by the asymptotic behaviour
\[
p_0(z)=z+\mathcal{O}\Big(\dfrac{1}{z}\Big),\quad
p_i(z)=u_i+\mathcal{O}\Big(\dfrac{1}{z}\Big),\, i=1,2;\quad \mbox{as $z\rightarrow\infty$}.
\]
There are four  critical points of $z(p)$ which give rise to four branch points $\pm x_1,\,\pm x_2$ in the $z$-plane where
\[
x_1=q_1\,\dfrac{\sqrt{1+8\,a^2}+3}{\sqrt{1+8\,a^2}+1},\quad
x_2=q_2\,\dfrac{\sqrt{1+8\,a^2}-3}{\sqrt{1+8\,a^2}-1},
\]
\[
q_{1,2}=\sqrt{\dfrac{1}{2}+a^2\pm\dfrac{1}{2}\sqrt{1+8\,a^2}}.
\]
\begin{center}
\begin{figure}[h!]
\centering
\includegraphics[width=6cm]{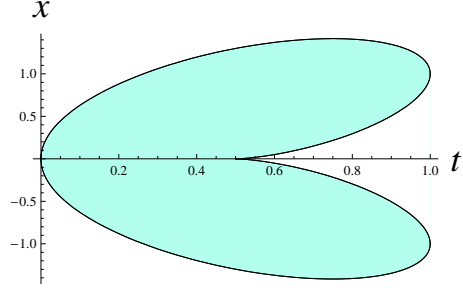}
\caption{Limit support for Brownian motions with two symmetric endpoints for $b=1$}
\end{figure}																						
\end{center}

It is easy to see that $x_1$ is real for all $a\geq 0$, while $x_2$ is real for $a\geq 1$ ($x_2<x_1$) and purely imaginary for $0<a<1$.
Now, from \eqref{n4h} and taking into account that $H(p)=p-z(p)$  we deduce that the eigenvalue density is given by
\begin{equation}\label{n4h2}
\rho_{0}(x)=\dfrac{1}{2\,i\pi}\,(H(p_{0+}(x))-H(p_{0-}(x))=
\dfrac{1}{2\,i\pi}\,(p_{0+}(x)-p_{0-}(x)),\quad x\in I_{0}.
\end{equation}
\begin{center}
\begin{figure}[h!]
\centering
\includegraphics[width=10cm]{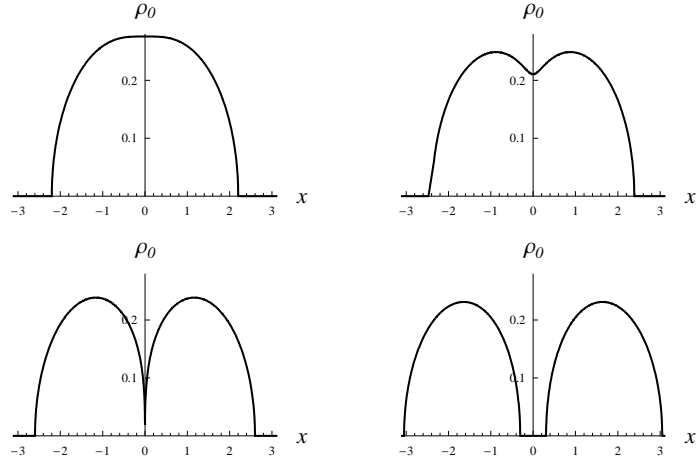}
\caption{The density  of Brownian motions  $\rho_0(x)$ for $a=1/2,\,3/4,\,1$ and $3/2$, respectively}
\end{figure}
\end{center}

Using Cardano's formula for $p_0$ one finds

\[
\rho_0(x)=\dfrac{2\,x^2+6\,(a^2-1)-\sqrt[3]{2}\,\Big( r(x)-\sqrt{r(x)^2-4\,s(x)^3} \Big)^{2/3}}{2^{5/3}\,\sqrt{3}\,\pi
\,\sqrt[3]{r(x)-\sqrt{r(x)^2-4\,s(x)^3}}},
\]
where
\[
r(x):=-2\,x^3+18\,a^2\,x+9\,x,\quad s(x):=x^2+3\,(a^2-1).
\]
The form of the support $I_0$ depends on the analytic properties of the function $p_0(z)$ ( see \cite{ortb1}-\cite{ortb4}):
\begin{description}
\item[a)] For $0<a\leq 1$ the function $p_0$ is analytic in $\C-[-x_1,x_1]$ and $I_0= [-x_1,x_1]$.

\item[b)] For $a> 1$ the function $p_0$ is analytic in $\C-(
[-x_1,-x_2]\cup[x_2,x_1])$ and $I_0= [-x_1,-x_2]\cup[x_2,x_1]$.
\end{description}

\newpage

\noindent {\bf Acknowledgements}

\vspace{0.3cm} The authors  wish to thank the  Spanish Ministerio de
Educaci\'on y Ciencia (research project FIS2008-00200/FIS) for its
finantial support. This work is also part of the MISGAM programme of
the European Science Foundation.

\end{document}